\documentclass[
preprint,
prd,tightenlines,showpacs,nofootinbib,
eqsecnum,amsfonts,amsmath,amssymb]{revtex4}

\usepackage{bm}
\usepackage{graphicx} 
\usepackage{hyperref}

\newcommand{\ui}{\mathrm{i}}
\newcommand{\ud}{\mathrm{d}}
\newcommand{\uD}{\mathrm{D}}
\newcommand{\calW}{\mathcal{W}}
\newcommand{\dcalW}{\calW_{{}_{\Pi_\perp}}}

\allowdisplaybreaks

\begin{document}

\title{Model of Dark Matter and Dark Energy \\ Based on Gravitational
Polarization}

\author{Luc \textsc{Blanchet}}\email{blanchet@iap.fr}
\affiliation{$\mathcal{G}\mathbb{R}\varepsilon{\mathbb{C}}\mathcal{O}$,
Institut d'Astrophysique de Paris --- UMR 7095 du CNRS, \\ Universit\'e Pierre
\& Marie Curie, 98\textsuperscript{bis} boulevard Arago, 75014 Paris, France}
\author{Alexandre \textsc{Le Tiec}}\email{letiec@iap.fr}
\affiliation{$\mathcal{G}\mathbb{R}\varepsilon{\mathbb{C}}\mathcal{O}$,
Institut d'Astrophysique de Paris --- UMR 7095 du CNRS, \\ Universit\'e Pierre
\& Marie Curie, 98\textsuperscript{bis} boulevard Arago, 75014 Paris, France}

\date{\today}

\begin{abstract}
A model of dark matter and dark energy based on the concept of gravitational
polarization is investigated. We propose an action in standard general
relativity for describing, at some effective or phenomenological level, the
dynamics of a dipolar medium, i.e. one endowed with a dipole moment vector, and
polarizable in a gravitational field. Using first-order cosmological
perturbations, we show that the dipolar fluid is undistinguishable from
standard dark energy (a cosmological constant $\Lambda$) plus standard dark
matter (a pressureless perfect fluid), and therefore benefits from the
successes of the $\Lambda$-CDM ($\Lambda$-cold dark matter) scenario at
cosmological scales. Invoking an argument of ``weak clusterisation''
of the mass distribution of dipole moments, we find that the dipolar
dark matter reproduces the phenomenology of the modified Newtonian
dynamics (MOND) at galactic scales. The dipolar medium action naturally
contains a cosmological constant, and we show that if the
model is to come from some fundamental underlying physics, the cosmological
constant $\Lambda$ should be of the order of $a_0^2/c^4$, where $a_0$ denotes
the MOND constant acceleration scale, in good agreement with observations.
\end{abstract}


\maketitle

\section{Introduction}\label{secI}

In the current concordance model of cosmology (the $\Lambda$-CDM scenario, see
e.g.~\cite{OsSt.95}) based on Einstein's general relativity (GR), the
mass-energy content of the Universe is made of roughly 4\% of baryons, 23\% of
cold dark matter (CDM) and 73\% of dark energy in the form of a cosmological
constant $\Lambda$. The dark matter accounts for the well-known discrepancy
between the mass of a typical cluster of galaxies as deduced from its
luminosity, and the Newtonian dynamical mass~\cite{Zw.33}. The model has so
far been very successful in reproducing the observed cosmic microwave
background (CMB) spectrum~\cite{HuDo.02} and explaining the distribution of
baryonic matter from galaxy clusters scale up to cosmological scales by the
non-linear growth of initial perturbations~\cite{Be.al.02}. Although the exact
nature of the hypothetical dark matter particle remains unknown,
super-symmetric extensions of the standard model of particle physics predict
well-motivated candidates (see~\cite{Be.al.05} for a review). Simulations
suggest some universal dark matter density profile around
galaxies~\cite{Na.al.97}. However, in that respect, the CDM hypothesis has some
difficulties~\cite{McBl.98,SaMc.02} at explaining in a natural way the
distribution and properties of dark matter at galactic scales.

The modified Newtonian dynamics (MOND) was proposed by
Milgrom~\cite{Mi1.83,Mi2.83,Mi3.83} to account for the basic phenomenology of
dark matter in galactic halos, as evidenced by the flat rotation curves of
galaxies, and the Tully-Fisher relation~\cite{TuFi.77} between the observed
luminosity and the asymptotic rotation velocity of spiral galaxies. However,
if MOND serves very well for these purposes (and some others
also~\cite{SaMc.02}), we know that MOND does not fully account for the
inferred dark matter at the intermediate scale of clusters of
galaxies~\cite{Ge.al.92,Sa.05,Cl.al.06}. In addition, MOND cannot be considered
as a viable physical model, but only as an \textit{ad-hoc} --- though
extremely useful --- phenomenological ``recipe''. In the usual interpretation,
MOND is viewed (see~\cite{Mi.02} for a review) as a modification of the
fundamental law of gravity or the fundamental law of dynamics, without the
need for dark matter. The relativistic extensions of MOND, of which the
Tensor-Vector-Scalar (TeVeS) theory~\cite{Sa.97,Be.04,Sa.05} is the prime
example, share this view of modifying the gravity sector, by postulating some
suplementary fields associated with the gravitational force, in addition to
the metric tensor field of GR (see~\cite{BrEs.07} for a review). Recently,
such modified gravity theories have evolved toward Einstein-{\ae}ther like
theories~\cite{JaMa.01,Ja.08,Zl.al.07,Zh.07,Ha.al.08}.

Each of these alternatives has proved to be very successful in complementary
domains of validity: the cosmological scale (and cluster scale) for the CDM
paradigm and the galactic scale for MOND. It is frustrating that two
successful models seem to be fundamentally incompatible. In the present paper
we shall propose a third approach, which has the potential of bringing
together the main aspects of both $\Lambda$-CDM and MOND in a single
relativistic model. Namely, we keep the standard law of gravity, i.e. GR and
its Newtonian limit, but we add to the distribution of ordinary matter some
specific non-standard form of dark matter (described by a relativistic action
in usual GR) in such a way as to naturally explain the phenomenology of MOND
at galactic scales. Furthermore, we prove that this form of dark matter leads
to the same predictions as for the $\Lambda$-CDM cosmological scenario at
large scales. In particular, we find that the relativistic action for this
matter model naturally contains the dark energy in the form of a cosmological
constant $\Lambda$. Thus, our model will benefit from both the successes of
the $\Lambda$-CDM scenario, and the MOND phenomenology.

The model will be based on the observation~\cite{Bl1.07,Bl2.07} that the
phenomenology of MOND can be naturally interpreted by an effect of
``gravitational polarization'' of some dipolar medium constituting the dark
matter. The effect can be essentially viewed (in a Newtonian-like
interpretation~\cite{Bl1.07}) as the gravitational analogue of the electric
polarization of a dielectric material, whose atoms can be modelled by electric
dipoles, in an applied electric field~\cite{Jac}. In the quasi-Newtonian model
of~\cite{Bl1.07} the gravitational polarization follows from a microscopic
description of the dipole moments in analogy with electrostatics. It was shown
that the gravitational dipole moments require the existence of some internal
non-gravitational force to stabilize them in a gravitational field. Thanks to
this internal force, an equilibrium state for the dipolar particle is
possible, in which the dipole moment is aligned with the gravitational field
and the medium is polarized. The MOND equation follows from that equilibrium
configuration. However the model~\cite{Bl1.07} cannot be considered as viable
because it is non-relativistic, and involves negative gravitational-type
masses (or gravitational charges) and consecutively a violation of the
equivalence principle at a fundamental level.

In a second model~\cite{Bl2.07} we showed that it is possible to describe
dipolar particles consistently with the equivalence principle by an action
principle in standard GR. The action depends on the particle's position in
space-time (as for an ordinary particle action) and also on a four-vector
dipole moment carried by the particle. The particle's position and the dipole
moment are considered to be two dynamical variables to be varied independently
in the action. Furthermore, a force internal to the dipolar particle was
introduced in the form of a scalar potential function (say $V$) in this
action. The potential $V$ depends on some adequately defined norm of the
dipole moment vector. Because of that force, the particle is not a ``test''
particle and its motion in space-time is non-geodesic. The non-relativistic
limit of the relativistic model~\cite{Bl2.07} was found to be different from
the quasi-Newtonian model~\cite{Bl1.07} (hence the two models are distinct)
but it was possible under some hypothesis to recover the same equilibrium
state yielding the MOND equation as in~\cite{Bl1.07}. However the relativistic
model~\cite{Bl2.07}, if considered as a model for dark matter, has some
drawbacks --- notably the mechanism of alignement of the dipole moment with
the gravitational field is unclear (so the precise link with MOND is
questionable), and the dynamics of the dipolar particles in the special case
of spherical symmetry does not seem to be very physical.

In the present paper, we shall propose a third model which will be based on an
action similar to that of the relativistic model~\cite{Bl2.07} but with
some crucial modifications. First we shall add, with respect to~\cite{Bl2.07},
an ordinary mass term in the action to represent the (inertial or passive
gravitational) mass of the dipolar particles. Second, the main improvement we
shall make is to assume that the internal force derives from a potential
function in the action (call it $\calW$) which depends not on the dipole
moment itself as in \cite{Bl2.07} but on the local \textit{density} of dipole
moments, i.e. the \textit{polarization} field. In this new approach we are
thus assuming that the motion of the dipolar particles is influenced by the
density of the surrounding medium. This is analogous to the description of a
plasma in electromagnetism in which the internal force, responsible for the
plasma oscillations, depends on the density of the plasma (cf. the expression
of the plasma frequency~\cite{Jac}).\footnote{In the quasi-Newtonian
model~\cite{Bl1.07} the dipolar medium was formulated as the gravitational
analogue of a plasma, oscillating at its natural plasma frequency.}
Because the action [given by \eqref{S} with \eqref{L} below] will now depend
on the density of the medium, it becomes more advantageous to write it as a
\textit{fluid} action rather than as a \textit{particle} action. 

This simple modification of the model, in which the potential $\calW$ depends
on the polarization field, will have important consequences. First of all, the
relation with the phenomenology of MOND will become clear and
straightforward. Secondly, we shall find that the motion of dipolar particles
in the central field of a spherical mass (in the non-relativistic limit) makes
now sense physically. The drawbacks of the previous model~\cite{Bl2.07} are
thus cured. Last but not least, we shall find that the model naturally
involves a cosmological constant. Then, with the equations of motion and
evolution (and stress-energy tensor) derived from the action, we show the
following:
\begin{enumerate}
\item The dipolar fluid is undistinguishable from standard dark energy (a
  cosmological constant) plus standard CDM (say a pressureless perfect fluid)
  at cosmological scales, i.e. at the level of \textit{first-order}
  cosmological perturbations.\footnote{Note however that while in the
  standard scenario the CDM particle is, say, a well-motivated supersymmetric
  particle (perhaps to be discovered at the LHC in CERN), in our case the fundamental
  nature of the ``dipolar particle'' will remain unknown.} The model is thus
  consistent with the observations of the CMB fluctuations. However, the model
  should differ from $\Lambda$-CDM at the level of second-order cosmological
  perturbations.
\item The MOND phenomenology of the flat rotation curves of galaxies and the
  Tully-Fisher relation is recovered at galactic scales (for a galaxy at low
  redshift) from the effect of gravitational polarization. There is a
  one-to-one correspondence between the MOND function (say $\mu=1+\chi$) and
  the potential function $\calW$ introduced in the action.
\item The minimum of the potential function $\calW$ is a cosmological constant
  $\Lambda$. We find that if $\calW$ is to be considered as ``fundamental'',
  i.e. coming from some fundamental underlying theory (presumably a quantum
  field theory), the cosmological constant should be numerically of the order
  of $a_0^2/c^4$, where $a_0$ denotes the MOND constant acceleration scale.
\end{enumerate}
A relation of the type $\Lambda \sim a_0^2/c^4$ between a cosmological
observable $\Lambda$ and a parameter $a_0$ measured from observations at
galactic scales is quite remarkable and is in good agreement with
observations. More precisely, if we define the natural acceleration scale
associated with the cosmological constant,
\begin{equation}\label{aLambda}
a_\Lambda = \frac{c^2}{2\pi}\,\sqrt{\frac{\Lambda}{3}}\, ,
\end{equation}
then the current astrophysical measurements yield $a_0 \simeq 1.3 \,
a_\Lambda$. The related numerical coincidence $a_0 \sim c H_0$ was pointed out
very early on by Milgrom \cite{Mi1.83,Mi2.83,Mi3.83}. The near agreement
between $a_0$ and $a_\Lambda$ has a natural explanation within our model,
although the exact numerical coefficient between the two acceleration scales
cannot be determined presently.

Since the present model will not be connected to any (quantum) fundamental
theory, it should be regarded merely as an ``effective'' or even
``phenomenological'' model. We shall even argue (though this remains open)
that it may apply only at large scales, from the galactic scale up to
cosmological scales, and not at smaller scales like in the Solar
System. However, this model offers a nice unification between the dark energy
in the form of $\Lambda$ and the dark matter in the form of MOND (both effects
of dark energy and dark matter occuring when gravity is weak). Furthermore, it
reconciles in some sense the observations of dark matter on cosmological
scales, where the evidence is for the standard CDM, and on galactic scales,
which is the realm of MOND. It would be interesting to study the intermediate
scale of clusters of galaxies and to see if the model is consistent with
observations. Such a study should probably be performed using numerical methods.

The plan of this paper is as follows. In section \ref{secII} we present the
action principle for the dipolar medium, and we vary the action to obtain the
equation of motion, the equation of evolution and the stress-energy tensor. In
section \ref{secIII} we apply first-order cosmological perturbations (on a
homogeneous and isotropic background) to prove that the dipolar fluid
reproduces all the features of the standard dark matter paradigm at
cosmological scales. We investigate the non-relativistic limit of the model in
section \ref{secIV}, and show that, under some hypothesis, the polarization of
the dipolar dark matter in the gravitational field of a galaxy results in an
apparent modification of the law of gravity in agreement with the MOND
paradigm. Section \ref{secV} summarizes and concludes the paper. The dynamics
of the dipolar dark matter in the central gravitational field of a spherically
symmetric mass distribution is investigated in appendix \ref{appA}.

\section{Dipolar fluid in general relativity}\label{secII}

\subsection{Action principle}\label{secIIA}

Our model will be based on a specific action functional for the dipolar fluid
in standard GR. This fluid is described by the four-vector current density
$J^\mu = \sigma u^\mu$, where $u^\mu$ is the four-velocity of the fluid,
normalized to $g_{\mu \nu} u^\mu u^\nu = -1$, and where $\sigma =
\sqrt{-g_{\mu \nu} J^\mu J^\nu}$ represents its rest mass
density.\footnote{Greek indices take the space-time values $\mu, \nu, \ldots =
0,1,2,3$ and Latin ones range on spatial values $i,j, \ldots = 1,2,3$. The
metric signature is $(-,+,+,+)$. The convention for the Riemann curvature
tensor $R^\mu_{\phantom{\mu} \nu \rho \sigma}$ is the same as in
\cite{MTW}. Symmetrization of indices is $(\mu\nu) \equiv \frac{1}{2}
(\mu\nu+\nu\mu)$ and $(ij) \equiv \frac{1}{2} (ij+ji)$. In sections
\ref{secII} and \ref{secIII} we make use of geometrical units $G = c = 1$.} In this
paper we shall conveniently rescale most of the variables used in
\cite{Bl2.07} by a factor of $2m$, where $m$ is the mass parameter introduced
in the action of~\cite{Bl2.07}. Hence we have $\sigma = 2m \, n$, where $n$ is
the number density of dipole moments in the notation of \cite{Bl2.07}. The
above current vector is conserved in the sense that
\begin{equation}\label{continuity}
	\nabla_\mu J^\mu = 0 \, ,
\end{equation}
where $\nabla_\mu$ denotes the covariant derivative associated with the metric
$g_{\mu \nu}$. Our fundamental assumption is that the dipolar fluid is endowed
with a dipole moment vector field $\xi^\mu$ which will be considered as a
dynamical variable. We have $\xi^\mu = \pi^\mu / 2m$ where $\pi^\mu$ is the
dipole moment variable used in~\cite{Bl2.07} (hence $\xi^\mu$ has the
dimension of a length).

Adopting a fluid description of the dipolar matter rather than a particle
formulation as in~\cite{Bl2.07},\footnote{The fluid action is obtained from
the particle one by the formal prescription $\sum \int \ud \tau \rightarrow
\int \ud^4 x \, \sqrt{-g} \,n$, where the sum runs over all the particles, and
$n$ is the number density of the fluid.} we postulate that the dynamics of the
dipolar fluid in a prescribed gravitational field $g_{\mu \nu}$ is derived
from an action of the type
\begin{equation}\label{S}
	S = \int \ud^4 x \, \sqrt{-g} \, L \bigl[ J^\mu, \xi^\mu,
	\dot{\xi}^\mu, g_{\mu \nu} \bigr] \, ,
\end{equation}
where $g = \text{det}(g_{\mu \nu})$, the integration being performed over the
entire 4-dimensional manifold. This action is to be added to the Einstein-Hilbert action for gravity, and to the actions of all the other matter fields. The Lagrangian $L$ depends on the current
density $J^\mu$, the dipole moment vector $\xi^\mu$, and its covariant
derivative $\dot{\xi}^\mu$ with respect to the proper time $\tau$ (such that
$\ud \tau = \sqrt{-g_{\mu \nu} \ud x^\mu \ud x^\nu}$), which is defined using
a fluid formulation by
\begin{equation}\label{xidot}
        \dot{\xi}^\mu \equiv \frac{\uD \xi^\mu}{\ud \tau} \equiv u^\nu
        \nabla_\nu \xi^\mu\, ,
\end{equation}
and where $\uD/\ud \tau$ is denoted by an overdot. In addition, the Lagrangian
depends explicitly on the metric $g_{\mu \nu}$ which serves at lowering and
raising indices, so that for instance $\dot{\xi}_\mu=g_{\mu \nu}\dot{\xi}^\nu$.

We shall consider an action for the dipolar medium similar to the one proposed
in~\cite{Bl2.07}, with however a crucial generalization in that the potential
function therein, which is supposed to describe a non-gravitational force
internal to the dipole moment, will be allowed to depend not only on the
dipole moment variable $\xi^\mu$, but also on the rest mass density of the
dipolar fluid $\sigma$. More precisely, we shall assume that the potential
function $\calW$ in the action depends on the dipole moment $\xi^\mu$ only
through the \textit{polarization}, namely the number density of dipole
moments, that is defined by
\begin{equation}\label{Pimu}
	\Pi^\mu = \sigma \xi^\mu \, ,
\end{equation}
or equivalently $\Pi^\mu = n \pi^\mu$ in the notation of~\cite{Bl2.07}. The
dynamics of dipolar particles will therefore be influenced by the local
density of the medium, in analogy with the physics of a plasma in which the
force responsible for the plasma oscillations depends on the density of the
plasma~\cite{Jac}. Our assumption is that $\calW$ is a function solely
of the norm $\Pi_\perp$ of the projection of the polarization field
\eqref{Pimu} perpendicular to the velocity, namely
\begin{equation}\label{Pi}
     \Pi_\perp = \sqrt{g_{\mu \nu} \Pi_\perp^\mu \Pi_\perp^\nu} =
     \sqrt{\perp_{\mu \nu} \! \Pi^\mu \Pi^\nu} \, .
\end{equation}
Here, the orthogonal projection of the polarization vector reads
$\Pi_\perp^\mu \! = \, \perp_\nu^\mu \! \Pi^\nu$, with the associated
projector defined by $\perp_{\mu \nu} \, \equiv g_{\mu \nu} + u_\mu
u_\nu$. Similarly, we can define $\xi_\perp^\mu = \, \perp_\nu^\mu \xi^\nu$
and its norm $\xi_\perp$ so that the (scalar) polarization field reads
\begin{equation}\label{Piexpr}
     \Pi_\perp=\sigma \xi_\perp\, .
\end{equation}
The chosen dependence of the internal potential on $\Pi_\perp$ will result in
important differences and improvements with respect to the model
of~\cite{Bl2.07}.

Our proposal for the Lagrangian of the dipolar fluid is
\begin{equation}\label{L}
	L = \sigma \left[ -1 - \sqrt{\bigl( u_\mu - \dot{\xi}_\mu \bigr)
	\bigl( u^\mu - \dot{\xi}^\mu \bigr)} + \frac{1}{2} \dot{\xi}_\mu \,
	\dot{\xi}^\mu \right] - \calW (\Pi_\perp) \, ,
\end{equation}
where the two dynamical fields are the conserved current vector $J^\mu =
\sigma u^\mu$ and the dipole moment vector $\xi^\mu$. The fourth term is our
fundamental potential which should in principle result from a more fundamental
theory valid at some microscopic level. The third term in \eqref{L} is the
same as in the previous model~\cite{Bl2.07} and clearly represents a
kinetic-like term for the evolution of the dipole moment vector. This term
will tell how this evolution should differ from parallel transport along the
fluid lines. The second term in \eqref{L} (also the same as in~\cite{Bl2.07})
is made of the norm of a space-like vector and is inspired by the known action
for the dynamics of particles with spin moving in a background gravitational
field~\cite{BaIs.80}. The motivation for postulating this term is that a
dipole moment can be seen as the ``lever arm'' of the spin considered as a
classical angular momentum (see a discussion in~\cite{Bl2.07}).

Finally, we comment on the first term in \eqref{L} which is a mass term in an
ordinary sense. The dipolar fluid we are considering will not be purely
dipolar (or mostly dipolar) as in the previous model~\cite{Bl2.07} but will
involve a monopolar contribution as well. Here we shall thus have some dark
matter in the ordinary sense. The mass term in \eqref{L} has been included for
cosmological considerations, so that we recover the ordinary dark matter
component at large scales (see section \ref{secIII}). However, one can argue
that the presence of such mass term $\sigma$ is not fine-tuned. Indeed, this
term corresponds to the simplest and most natural assumption that the relative
contributions of this mass density and the second and third terms in \eqref{L}
are comparable. In addition, we notice that $\sigma=2m\,n$ corresponds to the
inertial mass density of the dipole particles in the quasi-Newtonian
model~\cite{Bl1.07}, so it is natural by analogy with this model to include
that mass contribution in the action. Notice however that, even if the dipolar
fluid is endowed with a mass density in an ordinary sense, its dynamics is
well-defined only when the dipole moment is non-zero. Indeed, we observe that
the Lagrangian \eqref{L} becomes ill-defined when $\xi^\mu=0$ since the second
term in \eqref{L} is imaginary.

\subsection{Equations of motion and evolution}\label{secIIB}

In order to obtain the equations governing the dynamics of the dipolar fluid,
we vary the action \eqref{S} [with the explicit choice of the Lagrangian
\eqref{L}] with respect to the dynamical variables $\xi^\mu$ and $J^\mu$. The
calculation is very similar to the one performed in~\cite{Bl2.07}, but because
of the different notation adopted here for rescaled variables
(e.g. $\xi^\mu=\pi^\mu/2m$), and especially because of the more general form
of the potential function, we present all details of the derivation. Varying
first with respect to the dipole moment variable $\xi^\mu$, the resulting
Euler-Lagrange equation reads in general terms\footnote{We write the
Euler-Lagrange equation in this particle-looking form to emphasize the fact
that the action \eqref{L} is a \textit{particle} (or fluid) action. Of course,
this equation is equivalent to the usual \textit{field} equation $$ \nabla_\nu
\left( \frac{\partial L}{\partial \nabla_\nu \xi^\mu} \right) = \frac{\partial
L}{\partial \xi^\mu} \, .$$}
\begin{equation}\label{Lag1}
	\frac{\uD}{\ud \tau} \left( \frac{\partial L}{\partial \dot{\xi}^\mu}
	\right) + \nabla_\nu u^\nu \frac{\partial L}{\partial \dot{\xi}^\mu} =
	\frac{\partial L}{\partial \xi^\mu} \, ,
\end{equation}
in which the partial derivatives of the Lagrangian in \eqref{S} are applied
considering the four variables $\xi^\mu$, $\dot{\xi}^\mu$, $J^\mu$ and
$g_{\mu\nu}$ as independent. For the specific case of the Lagrangian
\eqref{L}, we get what shall be interpreted as the equation of \textit{motion}
of the dipolar fluid in the form
\begin{equation}\label{Kdot}
\dot{K}^\mu = - \mathcal{F}^\mu \, ,
\end{equation}
in which the left-hand-side (LHS) is the proper time derivative of the linear
momentum\footnote{The present notation is related to the one used
in~\cite{Bl2.07} by $K^\mu=P^\mu/2m$, $k^\mu=p^\mu/2m$,
$\mathcal{F}^\mu=F^\mu/m$ (and $\xi^\mu=\pi^\mu/2m$). The quantity called
$\Lambda$ in~\cite{Bl2.07} is now denoted $\Xi$ in order to avoid confusion
with the cosmological constant.}
\begin{equation}\label{K}
K^\mu = \dot{\xi}^\mu + k^\mu\, .
\end{equation}
Here, we introduced like in~\cite{Bl2.07} a special notation for a four-vector
$k^\mu$ which is \textit{space-like}, whose norm is normalized to $k^\mu k_\mu
= 1$, and which reads
\begin{equation}\label{kXi}
	k^\mu = \frac{u^\mu - \dot{\xi}^\mu}{\Xi} \, \quad \text{with} \quad
	\Xi = \sqrt{-1 - 2 u_\nu \dot{\xi}^\nu + \dot{\xi}_\nu \dot{\xi}^\nu}
	\, .
\end{equation}
The space-like four-vector $k^\mu$ will not represent the linear momentum (per
unit mass) of the particle --- that role will be taken by $K^\mu$ which, as we
shall see, will normally be time-like, see \eqref{Kexpr} below. The quantity
$\Xi$ has an important status in the present formalism because it represents
the second term in the Lagrangian \eqref{L} and we shall be able to set it to
one in section \ref{secIIC} as a particular way of selecting some physically
interesting solution. On the right-hand-side (RHS) of \eqref{Kdot}, the force
per unit mass acting on a dipolar fluid element is given by
\begin{equation}\label{Force}
	\mathcal{F}^\mu = \hat{\Pi}_\perp^\mu\,\dcalW\, ,
\end{equation}
in which we denote the unit direction of the polarization vector by
$\hat{\Pi}_\perp^\mu \equiv \Pi_\perp^\mu/\Pi_\perp = \xi_\perp^\mu/\xi_\perp$
and the ordinary derivative of the potential $\calW$ by $\dcalW \equiv \ud
\calW/\ud \Pi_\perp$. The ``\textit{internal}'' force \eqref{Force} being
proportional to the space-like four-vector $\xi_\perp^\mu = \,
\perp^\mu_\nu\xi^\nu$, we immediately get the constraint
\begin{equation}\label{uForce}
	u_\mu \, \mathcal{F}^\mu = 0\, .
\end{equation}

We now turn to the variation of the action with respect to the conserved
current $J^\mu = \sigma u^\mu$ (hence we deduce $\sigma=\sqrt{-J_\nu J^\nu}$
and $u^\mu=J^\mu/\sigma$). The general form of the Lagrange equation for the
conserved current density reads (see e.g.~\cite{Lic})\footnote{This can
alternatively be written with ordinary partial derivatives as
$$u^\nu \left[ \partial_\nu \left( \frac{\partial L}{\partial J^\mu} \right) -
\partial_\mu \left( \frac{\partial L}{\partial J^\nu} \right) \right] = 0\,
.$$}
\begin{equation}\label{Lag2}
	\frac{\uD}{\ud \tau} \left( \frac{\partial L}{\partial J^\mu} \right)
	= u^\nu \nabla_\mu \left( \frac{\partial L}{\partial J^\nu} \right)	.
\end{equation}
For the case at hands of the Lagrangian \eqref{L}, we get the following
equation, later to be interpreted as the \textit{evolution} equation for the
dipole moment,
\begin{equation}\label{Omegadot}
\dot{\Omega}^\mu = \frac{1}{\sigma}\nabla^\mu \left( \calW - \Pi_\perp \dcalW
\right) - R^\mu_{\phantom{\mu} \rho \nu \lambda} u^\rho \xi^\nu K^\lambda \, .
\end{equation}
A new type of linear momentum $\Omega^\mu$ --- having the same meaning as in
\cite{Bl2.07} --- has been introduced and defined by
\begin{equation}\label{Omega}
	\Omega^\mu = \omega^\mu - k^\mu \quad \text{with} \quad \omega^\mu =
	u^\mu \left( 1 + \frac{1}{2} \dot{\xi}_\nu \dot{\xi}^\nu + \xi_\perp
	\dcalW\right) - u_\nu \xi^\nu \mathcal{F}^\mu \, .
\end{equation}
The Riemann curvature term in the RHS of \eqref{Omegadot} represents the
analogue of the coupling to curvature in the Papapetrou equations of motion of
particles with spin in an arbitrary background~\cite{Pa.51}. The complete
dynamics and evolution of the dipolar fluid is now encoded into the equations
\eqref{Kdot} and \eqref{Omegadot}. Such equations constitute the appropriate
generalization for the case of a density-dependent potential $\calW$, and in
fluid formulation, of similar results in~\cite{Bl2.07}.

Notice that by contracting \eqref{Omegadot} with $J_\mu$, the second term in
the RHS of \eqref{Omegadot} cancels because of the symmetries of the Riemann
tensor, and we get
\begin{equation}\label{Jomegadot}
J_\mu \, \dot{\Omega}^\mu = \frac{\uD}{\ud\tau} \left( \calW - \Pi_\perp \dcalW
\right) .
\end{equation}
One can readily check that this constraint \eqref{Jomegadot} can alternatively
be derived from the other equation \eqref{Kdot} together with the definition
of $\Omega^\mu$ in \eqref{Omega}. On the other hand, contracting \eqref{Kdot}
with $u_\mu$ yields $u_\mu\dot{K}^\mu=0$, which according to the definition of
$K^\mu$, leads to the other constraint
\begin{equation}\label{contrainte2}
u_\mu \frac{\uD}{\ud\tau} \bigl[ \left( \Xi - 1 \right) k^\mu \bigr] = 0 \, .
\end{equation}
This constraint can be viewed as a differential equation for the variable
$\Xi$.

\subsection{Particular solution of the equations}\label{secIIC}

Following~\cite{Bl2.07}, we shall solve the constraint \eqref{contrainte2}
with the most obvious and natural choice of solution that
\begin{equation}\label{Xi1}
\Xi = 1\, .
\end{equation}
We shall see that this choice greatly simplifies the other equations we
have. In particular, we are going to prove that the equations of motion
\eqref{Kdot} and equations of evolution \eqref{Omegadot}, when reduced by the
condition $\Xi = 1$, finally depend only on the \textit{space-like} component
of the dipole moment that is orthogonal to the velocity, namely
$\xi_\perp^\mu$, so that the time-like component along the velocity,
i.e. $u_\nu \xi^\nu$, will have no physically observable consequences
(actually, in that case this unphysical component turns out to be complex
\cite{Bl2.07}).

The structure of the subsequent equations and the physical properties of the
model will heavily rely on the condition $\Xi = 1$. Note that we could regard
this condition not as a choice of solution but rather as a choice of
\textit{theory}. Indeed, we are going to pick up the simplest theory out of a
whole set of theories in which $\Xi$ could have some non trivial proper time
evolution obeying \eqref{contrainte2}. Actually, we can view the choice $\Xi =
1$ as an elegant way to impose into the Lagrangian formalism the condition
that \textit{in fine} the only physical component of the dipole moment should
be $\xi_\perp^\mu$, namely the one perpendicular to the four-velocity
field. We can imagine that it would be possible to impose the same physical
condition in a different way, for instance by using Lagrange multipliers into
the initial action. For exemple, in TeVeS~\cite{Sa.97,Be.04,Sa.05} or in
Einstein-{\ae}ther gravity~\cite{JaMa.01,Ja.08,Zl.al.07,Zh.07,Ha.al.08}, a
dynamical time-like vector field whose norm is unity is introduced by this
mean. However, the present situation is different because our final physical
vector $\xi_\perp^\mu$ is space-like.

When the condition \eqref{Xi1} holds, the two linear momenta \eqref{K} and
\eqref{Omega} simplify appreciably and we obtain
\begin{subequations}\label{KOmexpr}\begin{align}
	K^\mu &= u^\mu\, ,\label{Kexpr}\\ \Omega^\mu &= u^\mu \left( 1 +
	\xi_\perp \dcalW \right) + \perp^\mu_\nu \dot{\xi}_\perp^\nu \,
	.\label{Omexpr}
\end{align}\end{subequations}
We see that the linear momentum $K^\mu$ is finally time-like. These
expressions depend only on the orthogonal component $\xi_\perp^\nu$, and we
denote $\dot{\xi}_\perp^\nu \equiv \uD \xi_\perp^\nu / \ud \tau$. The
equations of motion and evolution take now the simple forms
\begin{subequations}\label{KOmdot}\begin{align}
\dot{u}^\mu &= - \mathcal{F}^\mu = - \hat{\Pi}_\perp^\mu\,\dcalW\,
,\label{Kdot_Xi}\\\dot{\Omega}^\mu &= \frac{1}{\sigma} \nabla^\mu \left( \calW
- \Pi_\perp \dcalW \right) - \xi_\perp^\nu R^\mu_{\phantom{\mu} \rho \nu
\lambda} u^\rho u^\lambda \, .\label{Omegadot_Xi}
\end{align}\end{subequations}
Finally, the whole dynamics of the dipolar fluid only depends on the space-like
perpendicular projection $\xi_\perp^\mu$ of the dipole moment.

\subsection{Expression of the stress-energy tensor}\label{secIID}

We vary the action \eqref{S} with respect to the metric $g_{\mu \nu}$ to
obtain the stress-energy tensor. We must first consider the general case where
$\Xi$ is unconstrained, and then only on the result make the restriction that
$\Xi=1$. We properly take into account the metric contributions coming from
the Christoffel symbols in the covariant time derivative $\dot{\xi}^\mu$ by
using the Palatini formula~\cite{Wei}. We are also careful that while the
dipole moment $\xi^\mu$ should be kept fixed during the variation, the
conserved current $J^\mu$ will vary because of the change in the volume
element $\sqrt{-g}\,\ud^4 x$. Instead of $J^\mu$, the relevant
metric-independent variable that has to be fixed is the ``coordinate'' current
density defined by $J_\ast^\mu = \sqrt{-g} \,J^\mu$. Straightforward
calculations yield the expression of the stress-energy tensor for an action of
the general type \eqref{S}. We find
\begin{align}\label{Tmunugen}
	T^{\mu \nu} &= 2 \frac{\partial L}{\partial g_{\mu \nu}} + g^{\mu \nu}
	\biggl( L - J^\rho \frac{\partial L}{\partial J^\rho} \biggr) + u^\mu
	u^\nu \, \dot{\xi}^\rho \frac{\partial L}{\partial \dot{\xi}^\rho}
	\nonumber\\ &+ \nabla_\rho \biggl( u^\mu u^\nu \frac{\partial
	L}{\partial \dot{\xi}_\rho} - u^\rho \xi^{(\mu} \frac{\partial
	L}{\partial \dot{\xi}_{\nu)}} - \xi^\rho u^{(\mu} \frac{\partial
	L}{\partial \dot{\xi}_{\nu)}} \biggr) \, ,
\end{align}
in which we denote $\partial L / \partial \dot{\xi}_\rho \equiv g^{\rho
\lambda} \partial L / \partial \dot{\xi}^\lambda$. The partial derivatives of
the Lagrangian are performed assuming that its ``natural'' arguments $J^\mu$,
$\xi^\mu$, $\dot{\xi}^\mu$ and $g_{\mu \nu}$ are independent. The application
to the particular case of the Lagrangian $\eqref{L}$ gives, for the moment for
a general value of $\Xi$,
\begin{equation}\label{Tmunu}
T^{\mu \nu} = - g^{\mu\nu} \left( \calW - \Pi_\perp \dcalW \right) +
\Omega^{(\mu} J^{\nu)} - \nabla_\rho \left( \left[ \xi^\rho K^{(\mu} - K^\rho
\xi^{(\mu} \right] J^{\nu)} \right) .
\end{equation}
In the second term of \eqref{Tmunu} we see that the linear momentum
$\Omega^\mu$ is related to the monopolar contribution to the stress-energy
tensor, while the other linear momentum $K^\mu$ parametrizes the dipolar
contribution in the third term. Comparing with equation (2.14) of
\cite{Bl2.07}, we observe that a new term, proportional to the metric
$g^{\mu\nu}$, has been introduced. This term will clearly be associated with a
cosmological constant, and we shall discuss it in detail below. One can
readily verify that the conservation law $\nabla_\nu T^{\mu \nu} = 0$ holds as
a consequence of the equation of conservation of matter \eqref{continuity},
and the equations of motion and evolution \eqref{Kdot} and \eqref{Omegadot},
for general $\Xi$.

In the next step we reduce the expression \eqref{Tmunu} by means of the
condition $\Xi=1$ and get
\begin{equation}\label{TmunuW}
	T^{\mu \nu} = - \calW \, g^{\mu\nu} + \sigma \left( u^\mu u^\nu +
	\xi_\perp \dcalW \! \perp^{\mu\nu} \! + \, u^{(\mu}\!\perp_\rho^{\nu)}
	\dot{\xi}_\perp^\rho \right) - \nabla_\rho \left( \left[
	\xi_\perp^\rho u^{(\mu} - u^\rho \xi_\perp^{(\mu} \right] J^{\nu)}
	\right) .
\end{equation}
Again we notice that this expression depends only on the perpendicular
projection $\xi_\perp^\mu$ of the dipole moment.

It will be useful in the following to decompose the stress-energy tensor
\eqref{TmunuW} according to the general canonical form
\begin{equation}\label{Tmunucan}
T^{\mu \nu} = r \, u^\mu u^\nu + \mathcal{P} \! \perp^{\mu \nu} + \, 2 \,
Q^{(\mu} u^{\nu)} + \Sigma^{\mu \nu} \, ,
\end{equation}
where $r$ and $\mathcal{P}$ represent the energy density and pressure, where
the ``heat flow'' $Q^\mu$ is orthogonal to the four-velocity, i.e. $u_\mu
Q^\mu = 0$, and the symmetric anisotropic stress tensor $\Sigma^{\mu \nu}$ is
orthogonal to the four-velocity and traceless, i.e. $u_\nu \Sigma^{\mu \nu} =
0$ and $\Sigma^\nu_{\nu} = 0$. We get
\begin{subequations}\label{rPQ}\begin{align}
r & = u_\rho u_\sigma T^{\rho \sigma}\, , \\ \mathcal{P} & = \frac{1}{3} \!
\perp_{\rho \sigma} \! T^{\rho \sigma}\, , \\ Q^\mu & = - \!
\perp^{\mu}_{\rho} u_\sigma \,T^{\rho \sigma} \, ,
\end{align}\end{subequations}
while the anisotropic stress tensor is obtained by subtraction. In the case
$\Xi=1$ where the dipolar fluid is described by the stress-energy tensor
\eqref{TmunuW} we find that the energy density, pressure, heat flow and
anisotropic stress tensor read respectively
\begin{subequations}\label{rPQS}\begin{align}
	r &= \calW - \Pi_\perp \dcalW \! + \rho \, , \label{r} \\
	\mathcal{P} &= - \calW + \frac{2}{3} \,\Pi_\perp
	\dcalW\, , \label{P}\\ Q^\mu &= \sigma\,\dot{\xi}_\perp^\mu +
	\Pi_\perp \dcalW u^\mu - \Pi_\perp^\lambda \nabla_\lambda u^\mu \, ,
	\label{Q}\\ \Sigma^{\mu \nu} &= \biggl( \frac{1}{3} \!  \perp^{\mu
	\nu} \! - \, \hat{\xi}_\perp^{\mu}\hat{\xi}_\perp^{\nu}\biggr)
	\Pi_\perp \dcalW
	\label{Sigmamunu} \, ,
\end{align}\end{subequations}
where we denote $\hat{\xi}_\perp^\mu \equiv \xi_\perp^\mu/\xi_\perp$, and
where we introduced for future use the convenient notation
\begin{equation}\label{gen_rho}
	\rho = \sigma - \nabla_\lambda \Pi_\perp^\lambda \, .
\end{equation}
By contrast to ordinary perfect fluids, the characteristic feature of the
dipolar fluid is the existence of non-vanishing heat flow $Q^\mu$ and
anisotropic stresses $\Sigma^{\mu\nu}$. Furthermore, we notice that the energy
density $r$ involves (\textit{via} $\rho$) a dipolar contribution given by
$-\nabla_\lambda \Pi_\perp^\lambda$. That contribution will play the crucial
role, as we will see in section \ref{secIV}, when recovering the phenomenology
of MOND.

\section{Cosmological perturbations at large scales}\label{secIII}

We are going to show in this section that the model of dipolar dark matter
[i.e. based on the action \eqref{S} and \eqref{L}, with equations of motion
reduced by the condition $\Xi=1$] contains the essential features of standard
dark matter at cosmological scales. We shall indeed prove that, at
\textit{first order} in cosmological perturbations, it behaves like a
pressureless perfect fluid. Furthermore, we shall see that the dipolar fluid
naturally contains a cosmological constant (the interpretation of which will
be discussed below), and is thus supported by the observations of dark
energy. The model is therefore consistent with cosmological observations of
the CMB fluctuations.

\subsection{Perturbation of the gravitational sector}\label{secIIIA}

We apply the theory of first-order cosmological perturbations around a
Friedman-Lema\^itre-Robertson-Walker (FLRW) background. For every generic
scalar field or component of a tensor field, say $F$, we shall write $F =
\overline{F} + \delta F$, where the background part $\overline{F}$ is the
value of $F$ in a FLRW metric, while $\delta F$ is a first-order perturbation
of this background value.

The FLRW metric is written in the usual way in terms of the conformal time
$\eta$, such that $\ud t = a \,\ud \eta$ where $a(\eta)$ is the scale factor
and $t$ the cosmic time, as
\begin{equation}
\ud \overline{s}^2 = \overline{g}_{\mu \nu} \, \ud x^\mu \ud x^\nu = a^2
\left[ - \ud \eta^2 + \gamma_{ij} \,\ud x^i \ud x^j \right] .
\end{equation}
Here $\gamma_{ij}$ is the metric of maximally symmetric spatial hypersurfaces
of constant curvature $K=0$ or $K=\pm 1$. The perturbed FLRW metric $\ud s^2 =
g_{\mu \nu} \, \ud x^\mu \ud x^\nu$ will be of the general form
\cite{PeUz}
\begin{equation}
\ud s^2 = a^2 \left[ - (1 + 2A) \,\ud \eta^2 + 2 \, h_i \, \ud \eta \, \ud x^i
+ (\gamma_{ij} + h_{ij}) \,\ud x^i \ud x^j \right] .
\end{equation}
Making use of the standard scalar-vector-tensor (SVT) decomposition
\cite{Ba.80,Mu.al.92}, the metric perturbations $h_i$ and $h_{ij}$ are
decomposed according to
\begin{subequations}\label{hhh}\begin{align}
h_i & = D_i B + B_i \, ,\\ h_{ij} & = 2 C \gamma_{ij} + 2 D_i D_j E + 2 D_{(i}
E_{j)} + 2 E_{ij} \, ,
\end{align}\end{subequations}
where $D_i$ denotes the covariant derivative with respect to the spatial
background metric $\gamma_{ij}$. The vectors $B^i$ and $E^i$ are
divergenceless, and the tensor $E^{ij}$ is at once divergenceless and
trace-free, i.e.
\begin{subequations}\label{condST}\begin{align}
D_i B^i = D_i E^i &= 0 \, , \\ D_i E^{ij} = E^i_i &= 0 \, .
\end{align}\end{subequations}
Spatial indices are lowered and raised with $\gamma_{ij}$ and its inverse
$\gamma^{jk}$. From these definitions, one can construct the gauge-invariant
perturbation variables
\begin{subequations}\label{PhiPsi}\begin{align}
	\Phi & \equiv A + (B' + \mathcal{H} B) - (E'' + \mathcal{H} E') \, ,\\
	\Psi & \equiv - C - \mathcal{H} (B - E') \, ,\\ \Phi_i & \equiv E_i' -
	B_i \, ,
\end{align}\end{subequations}
together with $E_{ij}$ which is already gauge-invariant. The prime stands for
a derivative with respect to the conformal time $\eta$, and $\mathcal{H}
\equiv a' / a$ denotes the conformal Hubble parameter. We shall also use the
alternative definition for a gauge-invariant gravitational potential
\begin{equation}
	X \equiv A - C - \left( \frac{C}{\mathcal{H}} \right)' = \Psi + \Phi +
	\left( \frac{\Psi}{\mathcal{H}} \right)' .
\end{equation}

\subsection{Kinematics of the dipolar fluid}\label{secIIIB}

The four-velocity of the dipolar fluid is decomposed into a background part
and a perturbation, $u^\mu = \overline{u}^\mu + \delta u^\mu$. We have both
$\overline{g}_{\mu\nu}\overline{u}^\mu\overline{u}^\nu=-1$ and
$g_{\mu\nu}u^\mu u^\nu=-1$. The background part is supposed to be comoving,
that is $\overline{u}^i=0$. This defines a zeroth order in the
perturbation. In a FLRW background this means that it will satisfy the
background geodesic equation $\dot{\overline{u}}^\mu = 0$. With standard
notations, we have
\begin{subequations}\label{udu}\begin{align}
	\overline{u}^\mu &= \frac{1}{a} \left( 1, \bm{0} \right) , \\ \delta
	u^\mu &= \frac{1}{a} \left( -A, \beta^i \right) ,
\end{align}\end{subequations}
while the covariant four-velocity will be written as $u_\mu = \overline{u}_\mu
+ \delta u_\mu$, with
\begin{subequations}\label{uducov}\begin{align}
	\overline{u}_\mu &= a \left( -1, \bm{0} \right) , \\ \delta
	u_\mu &= a \left( -A, \beta_i + h_i \right) .
\end{align}\end{subequations}
The velocities of all the other fluids (baryons, photons, neutrinos, \ldots)
are decomposed in a similar way. The perturbation of the three-velocity
$\beta^i$ is split into scalar and vector parts,
\begin{equation}\label{betai}
	\beta^i = D^i v + v^i \quad \text{with} \quad D_i v^i = 0 \, ,
\end{equation}
and we introduce the gauge-invariant variables describing the perturbed
motion,
\begin{subequations}\label{VVi}\begin{align}
V & \equiv v + E' \, ,\\ V_i & \equiv v_i + B_i \, .
\end{align}\end{subequations}

The dipolar dark matter fluid differs from standard dark matter by the
presence of the dipole moment $\xi_\perp^\mu$ (satisfying
$u_\mu\xi_\perp^\mu=0$) carried along the fluid trajectories. For the dipole
moment we also write a decomposition into a background part plus a
perturbation, namely $\xi_\perp^\mu = \overline{\xi}_\perp^\mu + \delta
\xi_\perp^\mu$. However, because a non-vanishing background dipole moment
would break the isotropy of space, and would therefore be incompatible with a
FLRW metric, we must make the assumption that the dipole moment is
\textit{zero} in the background, so that it is purely perturbative. Hence, we
pose
\begin{subequations}\label{xidxi}\begin{align}
	\overline{\xi}_\perp^\mu &= 0 \, , \\ \delta \xi_\perp^\mu &= \left(
	0, \lambda^i \right) ,
\end{align}\end{subequations}
where $\lambda^i$ represents the first-order perturbation of the dipole
moment. Beware of our notation for which $\lambda^i$ is a vector living in the
background spatial metric $\gamma_{ij}$. Thus the covariant components of the
dipole moment perturbation are ${\delta \xi_\perp}_\mu = \left( 0,
a^2\lambda_i \right)$ where $\lambda_i \equiv \gamma_{ij}\lambda^j$. Note that
there is no time component in the dipole moment perturbation because of the
constraint $u_\mu\xi_\perp^\mu=0$ which reduces to
$\overline{u}_\mu\delta{\xi}_\perp^\mu=0$ at linear order. Like for the
three-velocity field $\beta^i$ in \eqref{betai}, we split $\lambda^i$ into a
scalar and a vector part, namely
\begin{equation}\label{xii}
	\lambda^i = D^i y + y^i \quad \text{with} \quad D_i y^i = 0 \, .
\end{equation}
However, unlike for $v$ and $v^i$, we notice that $y$ and $y^i$ are
gauge-invariant perturbation variables. This is because the background
quantity is zero, $\overline{\xi}_\perp^\mu = 0$, hence the perturbation
$\delta \xi_\perp^\mu$ is gauge-invariant according to the Stewart-Walker
lemma~\cite{StWa.74,Ste}.

\subsection{Cosmological expansion of the fundamental potential}\label{secIIIC}

The next step is to make more specific our fundamental potential function
$\calW (\Pi_\perp)$ entering the Lagrangian \eqref{L}. Such function should be
a ``universal'' function of the polarization of the dipolar medium, described
by the polarization scalar field
\begin{equation}\label{Piexpr2}
\Pi_\perp = \sigma \xi_\perp\, .
\end{equation}
Now, we have seen that in cosmology there is no background (FLRW) value for
the dipole moment, hence the background value of the polarization field is
zero: $\overline{\Pi}_\perp=0$. In linear perturbations, the polarization is
expected to stay around the background value. Therefore, it seems physically
well-motivated that the value $\overline{\Pi}_\perp=0$ corresponds to a
minimum of the potential function $\calW$, so that $\Pi_\perp$ does not depart
too much from this background value, at least in the linear perturbation
regime. We therefore assume that $\calW (\Pi_\perp)$ is given
locally\footnote{The domain of validity of this expansion will be made more
precise in section \ref{secIVB}.} by an harmonic potential of the form
\begin{equation}\label{Wexp}
\calW (\Pi_\perp) = \calW_0 + \frac{1}{2}\calW_2 \, \Pi_\perp^2 +
\mathcal{O}\left(\Pi_\perp^3\right) ,
\end{equation}
where $\calW_0$ and $\calW_2$ are two constant parameters, and we pose
$\calW_1 = 0$. For linear perturbations, because $\Pi_\perp=\delta\Pi_\perp$
is already perturbative, we shall be able to neglect the higher order terms
$\mathcal{O}(\Pi_\perp^3)$ in \eqref{Wexp} because these will contribute to
second order at least in the internal force \eqref{Force}. Inserting the
\textit{ansatz} \eqref{Wexp} into \eqref{Force} we obtain
\begin{equation}\label{Fexp}
\mathcal{F}^\mu = \calW_2 \, \Pi_\perp^\mu +
\mathcal{O}\left(\Pi_\perp^2\right) .
\end{equation}
We asserted in the previous section that the background motion of the dipolar
fluid is geodesic, i.e. $\dot{\overline{u}}^\mu = 0$. This is now justified by
the fact that the force \eqref{Fexp} drives the non-geodesic motion
\textit{via} the equation of motion \eqref{Kdot_Xi}, hence since this force
vanishes in the background, the deviation from geodesic motion starts only at
perturbation order.

In the present model the coefficients $\calW_0$, $\calW_2$, \dots~of the
expansion of our fundamental potential $\calW (\Pi_\perp)$ are free
parameters, and therefore will have to be measured by cosmological or
astronomical observations. First of all, it is clear from inspection of the
action \eqref{L}, or from the general decomposition of the stress-energy
tensor [see \eqref{r} and \eqref{P}], that $\calW_0$ is nothing but a
cosmological constant, and we find
\begin{equation}\label{W0}
\calW_0 = \frac{\Lambda}{8\pi}\, .
\end{equation}
The coefficient $\calW_0$ is thereby determined by cosmological measurements
of ``dark energy''. As we shall show in section \ref{secIV}, the next two
coefficients $\calW_2$ and $\calW_3$ will be fixed by requiring that our model
reproduces the phenomenology of MOND at galactic scales~\cite{SaMc.02}, and
we shall find that $\calW_2 = 4 \pi$ and $\calW_3 = 32 \pi^2/a_0$ where $a_0$
is the constant MOND acceleration scale.

Hence, in this model the cosmological constant $\Lambda$ appears as the
minimum value of the potential function $\calW$, reached when the polarization
field is exactly zero, that is on an exact FLRW background (see
Fig.~\ref{mond}). Thus, it is tempting to interpret $\Lambda$ as a ``vacuum
polarization'', i.e. the residual polarization which remains when the
``classical'' part of the polarization $\Pi_\perp\rightarrow 0$. Of course our
model is only classical, hence there is no notion of vacuum polarization which
would be due to quantum fluctuations. However, we can imagine that the present
model is an effective one, describing at some macroscopic level a more
fundamental underlying quantum field theory (QFT) in which there is a
non-vanishing vacuum expectation value (VEV) of a quantum polarization field
giving rise to the observed cosmological constant~\cite{Ze.67}. Then, the
constant $\calW_0$ would play the role of the VEV of this hypothetical quantum
polarization field in such a more fundamental QFT.

\subsection{Perturbation of the dipolar fluid equations}\label{secIIID}

As for the four-velocity $u^\mu = \overline{u}^\mu + \delta u^\mu$, we
consider a linear perturbation of the rest mass energy density of the dipolar
fluid according to $\sigma = \overline{\sigma} + \delta \sigma$. The
conservation law $\nabla_\mu(\sigma u^\mu)=0$ reduces in the case of the
background to
\begin{equation}\label{sigmap}
\overline{\sigma}' + 3 \mathcal{H} \, \overline{\sigma} = 0 \, ,
\end{equation}
hence $\overline{\sigma}$ evolves like $a^{-3}$. Concerning the perturbation,
we define $\sigma \equiv \overline{\sigma} \left( 1 + \varepsilon \right)$ so
that the rest mass density contrast reads
\begin{equation}\label{eps}
\varepsilon = \frac{\delta \sigma}{\overline{\sigma}}\, .
\end{equation}
This quantity is not gauge-invariant, and one can associate with it in the
usual way a gauge-invariant variable by posing
\begin{equation}\label{epsF}
	\varepsilon_\text{F} \equiv \varepsilon -
	\frac{\overline{\sigma}'}{\overline{\sigma}} \frac{C}{\mathcal{H}} =
	\varepsilon + 3 C \, ,
\end{equation}
with the index F standing for ``flat slicing''. Alternatively, it is possible
to introduce other gauge-invariant variables, like for example
\begin{equation}\label{epsN}
     \varepsilon_\mathrm{N} \equiv \varepsilon - 3 \mathcal{H} ( B - E') =
     \varepsilon_\text{F} + 3 \Psi\, ,
\end{equation}
where the index N stands for ``Newtonian''.  For the linear perturbation, the
conservation law $\nabla_\mu(\sigma u^\mu)=0$ gives the gauge-invariant
equations
\begin{subequations}\label{epsilonprime}\begin{align}
	\varepsilon_\text{F}' + \Delta V &= 0 \,
	,\label{epsilonFprime}\\\varepsilon_\text{N}' + \Delta V &= 3 \Psi' \,
	,\label{epsilonNprime}
\end{align}\end{subequations}
where $\Delta = \gamma_{ij} D^i D^j$ denotes the usual Laplacian operator. In
the following we shall choose to work only with the flat-slicing variable
$\varepsilon_\text{F}$.

According to \eqref{Kdot_Xi}, the motion of the dipolar fluid obeys the
equation $\dot{u}^\mu = - \mathcal{F}^\mu$. A straightforward calculation
yields the gauge-invariant expression for the four-acceleration,
\begin{equation}\label{upoint}
	\dot{u}^\mu = \frac{1}{a^2}\Bigl( 0, D^i(\Phi + V' + \mathcal{H} V) +
	V^{i \, \prime} + \mathcal{H} V^i \Bigr)\, .
\end{equation}
On the other hand, the force is given by \eqref{Fexp} at first-order in the
perturbation, in which we can use
$\Pi^\mu_\perp=(0,\overline{\sigma}\lambda^i)$ to this order, with
$\lambda^i = D^i y + y^i$. Hence, in terms of gauge-invariant quantities, the
scalar and vector parts of the equation of motion read
\begin{subequations}\label{VViprime}\begin{align}
V' + \mathcal{H} \, V + \Phi &= - 4 \pi \, \overline{\sigma} \, a^2 \, y
\label{Vprime} \, ,\\ V_i' + \mathcal{H} \, V_i &= - 4 \pi \,
\overline{\sigma} \, a^2 \, y_i \label{Viprime} \, .
\end{align}\end{subequations}
Here we are anticipating on the results of the section \ref{secIV} and have
replaced the constant $\calW_2$ in the expression of the force \eqref{Fexp} by
its value $4 \pi$ determined from the comparison with MOND
predictions.

If there was no dipole moment (i.e. $y = y^i = 0$), we would recover the
standard geodesic equations for a perturbed pressureless perfect fluid (see
e.g.~\cite{PeUz}), and according to \eqref{Viprime}, the vector modes
would satisfy $(a V_i)'=0$, and therefore vanish like $a^{-1}$. In contrast
with the standard perfect fluid case, the dipolar fluid may have non-vanishing
vector modes because of the driving term proportional to $y_i$. Equation
\eqref{Vprime} clearly shows that the scalar modes are also affected by a
non-zero dipole moment.

The equation of evolution of the dipole moment was given by
\eqref{Omegadot_Xi}. Now, $\Omega^\mu$ reduces to $\dot{\xi}^\mu_\perp +
u^\mu$ at first perturbation order, hence the evolution equation gives at that
order
\begin{equation}
\ddot{\xi}^\mu_\perp + \dot{u}^\mu = - \xi^\nu_\perp
\overline{R}^\mu_{\phantom{\mu} \rho \nu \sigma} \overline{u}^\rho
\overline{u}^\sigma \, ,
\end{equation}
where $\overline{R}^\mu_{\phantom{\mu} \rho \nu \sigma}$ is the Riemann tensor
of the FLRW background. By easy calculations we find for the derivatives of
the dipole moment variable
\begin{subequations}\label{xiprime}\begin{align}
	\dot{\xi}_\perp^\mu &= \frac{1}{a} \Bigl( 0, \lambda^{i \, \prime} +
	\mathcal{H} \lambda^i \Bigr) \, , \\ \ddot{\xi}_\perp^\mu &=
	\frac{1}{a^2} \Bigl( 0, \lambda^{i \, \prime\prime} + \mathcal{H}
	\lambda^{i \, \prime} + \mathcal{H}' \lambda^i \Bigr) \, .
\end{align}\end{subequations}
The scalar and vector parts of the equation of evolution are thus given by
\begin{subequations}\label{yyiprime}\begin{align}
	y'' + \mathcal{H} \, y' &= - \left( V' + \mathcal{H} \, V + \Phi
	\right) \label{yprime} , \\ y_i'' + \mathcal{H} \,
	y_i' &= - \left( V_i' + \mathcal{H} \, V_i
	\right)
	\label{yiprime} .
\end{align}\end{subequations}
Notice that the equation for the vector modes can be integrated, giving the
simple relation
\begin{equation}\label{Vi_yi}
y_i' + V_i = \frac{s_i}{a} \, ,
\end{equation}
where $s_i$ is an integration constant three-vector.

A comment is in order at this stage. Recall that we have included in the
original Lagrangian \eqref{L} a mass term in the ordinary sense, with the most
natural value of the mass density simply given by $\sigma$. This choice was
made having in mind the physical analogy with the quasi-Newtonian model
\cite{Bl1.07} where $\sigma=2m\,n$ represented the inertial mass of the
dipolar particles. Now we can see on a more technical level that such mass
term is in fact essential for the model to work properly. If this mass term
was set to zero in the action, then the RHS of both equations \eqref{yprime}
and \eqref{yiprime} would be zero. We would then find that $y'$ and $y_i'$
vanish like $a^{-1}$, so that the dipole moment would in fact rapidly
disappear or at least become non-dynamical, and the whole model would turn out
to be meaningless.

Combining the equations of motion \eqref{VViprime} and the evolution equations
\eqref{yyiprime}, we obtain some differential equations for the scalar and
vector contributions $y$ and $y^i$ of the dipole moment $\lambda^i = D^i y +
y^i$, which turn out to be decoupled from the equations giving $V$ and $V^i$,
and to be exactly the same, \textit{viz}
\begin{subequations}\label{yyidecouple}\begin{align}
	y'' + \mathcal{H} \, y' - 4 \pi \, \overline{\sigma} \, a^2 \, y &= 0
	\label{y_evolu} \, ,\\ y_i'' + \mathcal{H} \, y_i' - 4 \pi \,
	\overline{\sigma} \, a^2 \, y_i &= 0
	\label{yi_evolu} \, .
\end{align}\end{subequations}
We find it remarkable that the dipole moment decouples from the other
perturbation variables so that its evolution depends \textit{in fine} only on
background quantities, namely $\overline{\sigma}$ and the scale factor $a$.
Since the equations for the scalar and vector modes are the same, we have also
the same equation for the dipole moment itself,
\begin{equation}\label{lambdai}
\lambda_i'' + \mathcal{H} \, \lambda_i' - 4 \pi \, \overline{\sigma} \, a^2 \,
\lambda_i = 0\, .
\end{equation}
Clearly, the solutions of \eqref{lambdai} behave typically as increasing and
decreasing exponentials moderated by a cosmologial damping term $\mathcal{H}
\, \lambda_i'$. We can also write this equation in terms of the cosmic time $t
= \int a \, \ud \eta$, namely\footnote{In this equation, the dot stands for a
derivative with respect to the coordinate time $t$, and not the proper time
$\tau$ as everywhere else.}
\begin{equation}\label{lambdai2}
    \ddot{\lambda}_i + 2 H \, \dot{\lambda}_i - 4 \pi \, \overline{\sigma} \,
	\lambda_i = 0\, ,
\end{equation}
where $H \equiv \dot{a}/a = a'$ is the usual Hubble parameter. We find that
the equations \eqref{lambdai} or \eqref{lambdai2} are the same as the equation
governing the growth of the density contrast of a perfect fluid with vanishing
pressure for the sub-Hubble modes (say $k\gg H$) and when we neglect the
contribution of other fluids; see \eqref{evoldeltaF2} below. In particular
this means that like for the case of the density of a perfect fluid there is
no problem of divergence (i.e. blowing up) of the components of the dipole
moment $\lambda_i$ between, say, the end of the inflationary era and the
recombination. We can thus apply the theory of first-order cosmological
perturbations even for the components of the dipole moment itself, which
should stay perturbative.

Notice that the value of the coefficient $\calW_2=4\pi$ used in
\eqref{lambdai} or \eqref{lambdai2}, which makes such equations identical with
the equation of growth of cosmological structures in the standard CDM
scenario, will only be determined in section \ref{secIV} from a comparison
with MOND predictions. There is thus an interesting interplay between the
cosmology at large scales and the gravitational physics of smaller
scales.\footnote{Actually the coefficient $4\pi$ in \eqref{lambdai} could be
changed if we had assumed a mass term in the action \eqref{L} different from
$\sigma$ (say $2\sigma$ or $\sigma/2$). The simplest choice we have made (for
different reasons) that $\sigma$ is the correct mass term in the action
corresponds also to the usual-looking evolution equation \eqref{lambdai}.}

\subsection{The perturbed stress-energy tensor}\label{secIIIE}

Consider next the stress-energy tensor of the dipolar fluid, that we
decomposed as \eqref{Tmunucan} with the expressions
\eqref{rPQS}--\eqref{gen_rho}. At first perturbation order, these expressions
reduce to
\begin{subequations}\label{rPQSlin}\begin{align}
	r &= \calW_0 + \rho\, ,\\ \mathcal{P} &= - \calW_0 \, ,\\ Q^\mu &=
	\frac{1}{a} \Bigl( 0, \overline{\sigma} \lambda^{i \, \prime} \Bigr)
	\, ,\label{Qmu} \\ \Sigma^{\mu \nu} &= 0 \, ,
\end{align}\end{subequations}
together with
\begin{equation}\label{rho0}
	\rho = \overline{\sigma} \,\bigl( 1 + \varepsilon - D_i \lambda^i
	\bigr)\, .
\end{equation}
We first note that part of the dipolar medium is actually made of a fluid of
``dark energy'' satisfying $\rho_\text{de} = - P_\text{de} = \calW_0 = \Lambda
/ 8 \pi$ where $\Lambda$ is the cosmological constant. Accordingly, we shall
write the decomposition
\begin{equation}\label{decompdedm}
	T^{\mu \nu} = T_\text{de}^{\mu \nu} + T_\text{dm}^{\mu \nu}\, ,
\end{equation}
where the stress-energy tensor associated with the cosmological constant is
denoted by $T_\text{de}^{\mu \nu}$, and where the other part represents
specifically a fluid of ``dark matter'' whose stress-energy tensor is
$T_\text{dm}^{\mu \nu}$. Their explicit expressions read
\begin{subequations}\label{Tmunudedm}\begin{align}
	T_\text{de}^{\mu \nu} &= - \calW_0 \, g^{\mu \nu}\,
	,\label{Tmunude}\\T_\text{dm}^{\mu \nu} &= \rho \, u^\mu u^\nu + 2 \,
	Q^{(\mu} u^{\nu)}\, .\label{Tmunudm}
\end{align}\end{subequations}
Note that the dark matter part of the dipolar fluid, which may be called
dipolar dark matter, has no pressure $P$, no anisotropic stresses $\Sigma^{\mu
\nu}$, but a heat flow $Q^\mu$ given by \eqref{Qmu} and an energy density
$\rho$ given by \eqref{rho0}, or alternatively
\begin{equation}\label{rho}
     \rho = \overline{\sigma} \left( 1 + \varepsilon - \Delta y \right) .
\end{equation}
The background energy density is simply given by the background rest mass
energy density, $\overline{\rho} = \overline{\sigma}$, and the corresponding
energy density contrast is
\begin{equation}\label{contrastdelta}
	\delta \equiv \frac{\delta \rho}{\overline{\rho}} = \varepsilon -
	\Delta y\, .
\end{equation}
It differs from the rest mass energy density contrast $\varepsilon$ because of the
internal dipolar energy. Like for $\varepsilon$, one can construct several
gauge-invariant perturbations associated with $\delta$. We shall limit
ourselves to the flat-slicing (F) one defined by (recall that $y$ is gauge-invariant)
\begin{equation}\label{deltaF}
    \delta_\text{F} \equiv \delta + 3 C = \varepsilon_\text{F} - \Delta y \, ,
\end{equation}
and whose evolution equation is
\begin{equation}\label{evoldeltaF}
	\delta_\text{F}' + \Delta V + \Delta y' = 0\, .
\end{equation}
Similar gauge-invariant density contrast variables are also defined for the
other fluids. Next, we split the dark matter stress-energy tensor
\eqref{Tmunudm} into a background part plus a linear perturbation, namely
$T_\text{dm}^{\mu \nu} = \overline{T}_\text{dm}^{\mu \nu} + \delta
T_\text{dm}^{\mu \nu}$, and find
\begin{subequations}\label{split}\begin{align}
	\overline{T}_\text{dm}^{\mu \nu} &= \overline{\rho} \,
	\overline{u}^\mu \overline{u}^\nu \, , \label{split0}\\ \delta
	T_\text{dm}^{\mu \nu} &= \delta \rho \, \overline{u}^\mu
	\overline{u}^\nu + 2 \, \overline{\rho} \, \delta u^{(\mu} \,
	\overline{u}^{\nu)} + 2 \, Q^{(\mu} \, \overline{u}^{\nu)} \,
	.\label{split1}
\end{align}\end{subequations}
We made use of the fact that the heat flow $Q^\mu$ is already perturbative to
replace the four-velocity in the last term by its background value.

We are now going to show that the dipolar dark matter stress-energy tensor is
undistinguishable, at linear perturbation order, from that of a perfect fluid
with vanishing pressure. To this end, we introduce the effective perturbed
four-velocity
\begin{equation}\label{u_eff}
	\delta \widetilde{u}^\mu \equiv \delta u^\mu +
	\frac{Q^\mu}{\overline{\rho}} \, .
\end{equation}
Notice that $\widetilde{u}^\mu=\overline{u}^\mu+\delta\widetilde{u}^\mu$ is
still an admissible velocity field because $\delta \widetilde{u}^0 = -A/a$ by
virtue of the transversality property $\overline{u}_\mu Q^\mu=0$. The
perturbed part of the dark matter stress-energy tensor \eqref{split1} can then
be written in the form
\begin{equation}
	\delta T_\text{dm}^{\mu \nu} = \delta \rho \, \overline{u}^\mu
	\overline{u}^\nu + 2 \, \overline{\rho} \, \delta \widetilde{u}^{(\mu}
	\, \overline{u}^{\nu)} \, ,
\end{equation}
which, together with \eqref{split0}, is precisely the stress-energy tensor of
a perfect fluid with vanishing pressure $P$, vanishing anisotropic stresses
$\Sigma^{\mu \nu}$, and a four-velocity field
$\widetilde{u}^\mu=\overline{u}^\mu+\delta\widetilde{u}^\mu$. Using the
definition \eqref{u_eff} of the perturbed four-velocity $\delta
\widetilde{u}^\mu$, with the explicit expression of the heat flow \eqref{Qmu},
one can check that this perfect fluid consistently follows a geodesic motion,
i.e. $\delta \dot{\widetilde{u}}^\mu = 0$.

\begin{figure}
	\includegraphics[width=8cm]{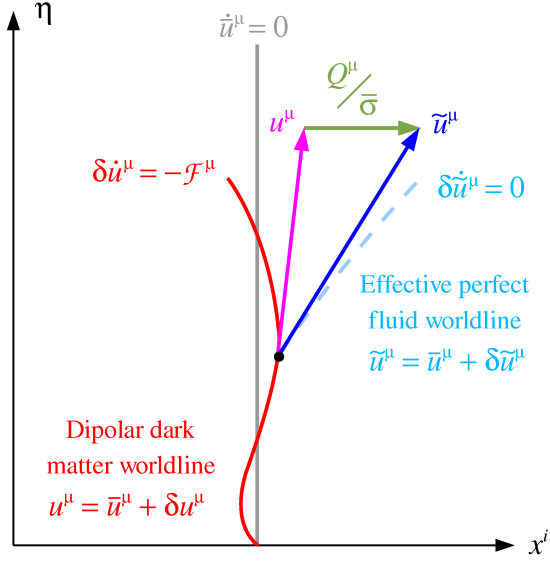}
	\caption{\footnotesize Sketch of the equivalence at \textit{first
	order} of cosmological perturbations between dipolar dark matter and
	an effective perfect fluid. The dipolar dark matter has a
	four-velocity $u^\mu = \overline{u}^\mu + \delta u^\mu$, and follows a
	non-geodesic motion driven by the internal force $\mathcal{F}^\mu$,
	namely $\dot{u}^\mu = - \mathcal{F}^\mu$. One can construct from
	$u^\mu$ and the heat flux $Q^\mu$ an effective four-velocity
	$\widetilde{u}^\mu = \overline{u}^\mu + \delta \widetilde{u}^\mu$
	satisfying a geodesic motion, i.e. $\dot{\widetilde{u}}^\mu = 0$. This
	is the four-velocity field of the effective perfect fluid associated
	with dipolar dark matter.}
	\label{ddm_pf}
\end{figure}

More explicitly, we can write the latter effective perturbation of the
four-velocity in the standard form $\delta \widetilde{u}^\mu = a^{-1}( -A,
\widetilde{\beta}^i)$, and find that the effective ordinary velocity reads
\begin{equation}\label{betaeff}
       \widetilde{\beta}^i = \beta^i + \lambda^{i\,\prime}\, ,
\end{equation}
which can be viewed as a modification of the space-like component of the
dipolar dark matter four-velocity. This allows one to build a new
four-velocity which would be tangent to the worldline of the effective perfect
fluid (cf. Fig.~\ref{ddm_pf}).
In terms of scalar and vector parts, if we
write $\widetilde{\beta}_i = D_i \widetilde{v} + \widetilde{v}_i$, then
\begin{subequations}\label{vvitilde}\begin{align}
	\widetilde{v} & = v + y' \, , \\ \widetilde{v}_i & = v_i + y_i' \, .
\end{align}\end{subequations}
Like for the perturbed four-velocity $\delta u^\mu$, we can introduce the
gauge-invariant variables
\begin{subequations}\label{VVieff}\begin{align}
	\widetilde{V} & \equiv \widetilde{v} + E' = V + y' \, , \\
	\widetilde{V}_i & \equiv \widetilde{v}_i + B_i = V_i + y_i' \, .
\end{align}\end{subequations}
In terms of the gauge-invariant variables $\widetilde{V}$, $\widetilde{V}_i$
and $\delta_\text{F}$, the dipolar dark matter fluid equations
\eqref{VViprime} and \eqref{evoldeltaF} finally read
\begin{subequations}\label{effeqns}\begin{align}
	\widetilde{V}' + \mathcal{H} \, \widetilde{V} + \Phi & = 0 \, ,
	\label{Vtilde'} \\ \widetilde{V}_i' + \mathcal{H} \, \widetilde{V}_i &
	= 0 \, , \label{Vitilde'} \\ \delta_\text{F}' + \Delta \widetilde{V} &
	= 0 \label{deltaF'} \, .
\end{align}\end{subequations}
These are precisely the standard evolution equations of a perfect fluid with
no pressure and no anisotropic stresses (see e.g. \cite{PeUz}).

To summarize, we have proved that at \textit{first order} of perturbation
theory --- and only at that order --- the dipolar fluid behaves exactly as
ordinary dark energy (i.e. a cosmological constant) plus ordinary dark matter
(i.e. a perfect fluid). If we specify the background rest mass energy density
$\overline{\sigma}$ so that $\Omega_\text{dm} \equiv 8 \pi \overline{\sigma}_0
/ 3 H_0^2 \simeq 0.23$ today as evidenced in cosmological observations, we can
assert that the first-order cosmological perturbation theory with the dipolar
fluid described by the stress-energy tensor
\eqref{decompdedm}--\eqref{Tmunudedm} will lead to the same predictions than
the standard $\Lambda$-CDM scenario --- and is therefore consistent with
cosmological observations at large scales. However, at \textit{second order}
of cosmological perturbations, the dipole moment entering the stress-energy
tensor cannot be absorbed in an effective perturbed velocity field, which
means that the dipolar dark matter fluid could in principle be distinguished
from a standard perturbed dark matter fluid. Working out the theory of
second-order cosmological perturbations could thus yield distinctive features
of the present model and reveal a signature of the dipolar nature of dark
matter. We have particularly in mind effects linked with the non-gaussianity
of the CMB fluctuations that are associated with second-order perturbations.

\subsection{Perturbation of the Einstein equations}\label{secIIIF}

The Einstein equations at first perturbation order around the FLRW background
read
\begin{equation}\label{EE}
	\delta G^{\mu \nu} = 8 \pi \Bigl( \delta T^{\mu \nu} + \sum_\text{f}
	\delta T_\text{f}^{\mu \nu} \Bigr) \, ,
\end{equation}
where $G^{\mu \nu}\equiv R^{\mu \nu}-\frac{1}{2}g^{\mu \nu}R$ is the Einstein
tensor and where $\delta T^{\mu \nu}=\delta T^{\mu \nu}_\text{de}+\delta
T_\text{dm}^{\mu \nu}$ is the perturbative part of the stress-energy tensor of
the dipolar fluid given by \eqref{Tmunudedm}. The summation runs over all the
other cosmological fluids present (baryons, photons, neutrinos, \ldots) which
are described by stress-energy tensors $T^{\mu \nu}_\text{f}$. Separating out
the dark matter from the dark energy (using the link $\calW_0=\Lambda/8\pi$)
we get
\begin{equation}\label{EELambda}
	\delta G^{\mu \nu} + \Lambda\,\delta g^{\mu \nu} = 8\pi \Bigl( \delta
	T_\text{dm}^{\mu \nu} + \sum_\text{f} \delta T_\text{f}^{\mu \nu}
	\Bigr) \, .
\end{equation}
As we have seen in the previous section, the dark matter fluid is entirely
described at linear perturbation order by the gauge-invariant variables
$\widetilde{V}$, $\widetilde{V}_i$ and $\delta_\text{F}$ (and the background
density $\overline{\rho}$) obeying the evolution equations \eqref{effeqns}
like for an ordinary pressureless fluid. We can thus immediately write the
gauge-invariant perturbation equations in the standard SVT formalism (see
e.g.~\cite{PeUz}). Though these are well-known, we reproduce them here for
completeness. For the scalar modes, we have
\begin{subequations}\label{Smode}\begin{align}
	\Delta \Psi - 3 \mathcal{H}^2 X & = 4\pi \,a^2 \Bigl( \overline{\rho}
	\, \delta_\text{F} + \sum_\text{f} \overline{\rho}_\text{f} \,
	\delta_\text{f}^\text{F} \Bigr)
	\label{Psi} \, ,\\ \Psi - \Phi & = 8\pi \,a^2 \sum_\text{f}
	\overline{\rho}_\text{f} \, w_\text{f} \, \sigma_\text{f}
	\label{PsiPhi} \, ,\\ \Psi' + \mathcal{H} \, \Phi & = - 4\pi \,a^2 \Bigl(
	\overline{\rho} \, \widetilde{V} + \sum_\text{f}
	\overline{\rho}_\text{f} \left( 1 + w_\text{f} \right) V_\text{f}
	\Bigr) \, ,\\ \mathcal{H} \, X' + \left( \mathcal{H}^2 + 2
	\mathcal{H}' \right) X & = 4\pi \,a^2 \sum_\text{f}
	\overline{\rho}_\text{f} \Bigl( w_\text{f} \, \Gamma_\text{f} +
	c_\text{f}^2 \, \delta_\text{f}^\text{F} + \frac{2}{3} w_\text{f} \,
	\Delta \sigma_\text{f} \Bigr) \, ,
\end{align}\end{subequations}
where we have singled out the contribution of the dipolar dark matter (cf. the
variables $\widetilde{V}$, $\delta_\text{F}$ and $\overline{\rho}$) from the other fluid
contributions described by their background density
$\overline{\rho}_\text{f}$, equation of state $w_\text{f}$, adiabatic sound
velocity $c_\text{f}$, and gauge-invariant entropy perturbation
$\Gamma_\text{f}$. We also introduced the SVT components of the perturbative
part of the anisotropic stress tensor, defined by $\delta \Sigma_\text{f}^{ij}
= a^2 \, \overline{\rho}_\text{f} \, w_\text{f} \, \bigl[ \Delta^{ij}
\sigma_\text{f} + D^{(i} \sigma_\text{f}^{j)} + \sigma_\text{f}^{ij} \bigr]$
with $\Delta^{ij} \equiv D^i D^j - \gamma^{ij} \Delta /3$. The variables
$\sigma_\text{f}$, $\sigma_\text{f}^i$ and $\sigma_\text{f}^{ij}$ are
gauge-invariant because the background part of $\Sigma_\text{f}^{ij}$ vanishes
in the case of a perfect fluid. The equations for the vector and tensor modes are
\begin{subequations}\label{VTmode}\begin{align}
	\left( \Delta + 2 K \right) \Phi^i & = - 16\pi \,a^2 \Bigl(
	\overline{\rho} \, \widetilde{V}^i + \sum_\text{f}
	\overline{\rho}_\text{f} \left( 1 + w_\text{f} \right) V_\text{f}^i
	\Bigr) \, ,\\ \Phi^{i\,\prime} + 2 \mathcal{H} \,\Phi^i & = 8\pi \,a^2
	\sum_\text{f} \overline{\rho}_\text{f} \, w_\text{f} \,
	\sigma_\text{f}^i \, ,\\E^{ij \,\prime\prime} + 2 \mathcal{H} \,E^{ij
	\,\prime} + (2K - \Delta) E^{ij} &= 8\pi \,a^2 \sum_\text{f}
	\overline{\rho}_\text{f} \, w_\text{f} \, \sigma_\text{f}^{ij} \, .
\end{align}\end{subequations}
We highlight once more the fact that at first perturbation order, the dipolar
dark matter is like ordinary dark matter, as can be seen from the fluid equations
\eqref{effeqns} and the Einstein equations
\eqref{Smode}--\eqref{VTmode}. Indeed, these sets of equations can be evolved
without any reference to the dipole moment $\lambda^i$. 

Combining the dipolar dark matter equations \eqref{Vtilde'} and
\eqref{deltaF'} with the Einstein equations \eqref{Psi}--\eqref{PsiPhi}, we
get the equation governing the growth of the dipolar dark matter density
contrast as
\begin{equation}\label{evoldeltaF2}
	\delta_\text{F}'' + \mathcal{H} \, \delta_\text{F}' - 4 \pi \,
	\overline{\rho} \, a^2 \, \delta_\text{F} = 3 \mathcal{H}^2 X + 4\pi
	\,a^2 \sum_\text{f} \overline{\rho}_\text{f} \,
	\Bigl(\delta_\text{f}^\text{F} - 2 w_\text{f} \, \Delta
	\sigma_\text{f} \Bigr)\, .
\end{equation}
Again, we find that the growth of structures driven by the equation
\eqref{deltaF'} or equivalently \eqref{evoldeltaF2} for the dipolar dark
matter of the present model is identical with that in the standard CDM model
at linear perturbation order. For sub-Hubble modes one can neglect the first
term in the RHS, and we expect that the contribution of the dark matter
dominates that of the other fluids, so we can neglect also the second term in
the RHS of \eqref{evoldeltaF2}.

Interestingly, we have found in \eqref{lambdai} that each of the components of
the dipole moment obey the same equation as \eqref{evoldeltaF2} but with
exactly zero RHS. Recall that the dipolar dark matter density contrast is
defined by \eqref{deltaF} as
\begin{equation}\label{deltaFdef}
	\delta_\text{F} = \varepsilon_\text{F} - D^i\lambda_i\, .
\end{equation}
From \eqref{lambdai} we see that the internal energy due to the dipole moment
satisfies the ``homogeneous'' equation that is associated with
\eqref{evoldeltaF2}, \textit{viz.} (recalling $\overline{\rho} = \overline{\sigma}$)
\begin{equation}\label{Dilambdai}
D^i\lambda_i'' + \mathcal{H} \, D^i\lambda_i' - 4 \pi \, \overline{\rho} \,
a^2 \, D^i\lambda_i = 0\, .
\end{equation}
This result indicates that, in the \textit{non-linear} regime, the
internal energy related to the dipole moment may contribute significatively to
the growth of perturbations (see section \ref{secIVB} for more comments).
Finally, it is clear that the rest-mass density
contrast obeys the same ``inhomogeneous'' equation, i.e.
\begin{equation}\label{evolepsF}
	\varepsilon_\text{F}'' + \mathcal{H} \, \varepsilon_\text{F}' - 4 \pi
	\, \overline{\rho} \, a^2 \, \varepsilon_\text{F} = 3 \mathcal{H}^2 X
	+ 4\pi \,a^2 \sum_\text{f} \overline{\rho}_\text{f} \,
	\Bigl(\delta_\text{f}^\text{F} - 2 w_\text{f} \, \Delta
	\sigma_\text{f} \Bigr)\, .
\end{equation}

\section{Dipolar dark matter at galactic scales}\label{secIV}

In this section, we shall show that, under some well motivated assumptions,
the dipolar dark matter naturally recovers the phenomenology of MOND for a
typical galaxy at low redshift. Such a link between a form of dipolar dark
matter and MOND was the primary motivation of previous
works~\cite{Bl1.07,Bl2.07}. We shall see that with the present improvement of
the model with respect to~\cite{Bl2.07}, thanks to the fact that the
fundamental potential in the action now depends on the polarization field
$\Pi_\perp=\sigma\xi_\perp$ (instead of $\xi_\perp$ in the previous
model~\cite{Bl2.07}), the relation with MOND is straightforward and physically
appealing.

\subsection{Non-relativistic limit of the model}\label{secIVA}

We investigate the non-relativistic (NR) limit of the dipolar fluid dynamics
described by the equations \eqref{Kdot_Xi} and \eqref{Omegadot_Xi}, and by the
stress-energy tensor \eqref{TmunuW}. To do so, we consider the
formal limit $c \rightarrow +\infty$,\footnote{From now on, we reintroduce for
convenience all factors of $c$ and $G$.} which is equivalent to the condition
$v \ll c$, where $v$ is the typical value of the coordinate three-velocity of
the dipolar fluid. To consistently keep track of the order of relativistic
corrections, we systematically write as $\mathcal{O} \left( c^{-n} \right)$ a
typical neglected remainder.

We are interested in the dynamics of dipolar dark matter and ordinary baryonic
matter in a typical galaxy at low redshift. Let us introduce a \textit{local}
Cartesian coordinate system $\{ ct, z^i \}$, centered on this galaxy around
some cosmological epoch, and which is \textit{inertial} in the sense that it
is without any rotation, nor acceleration with respect to some averaged
cosmological matter distribution at large distances from the galaxy. Such a
local coordinate system can be derived from the cosmological coordinate system
$\{ \eta, x^i \}$ used in section \ref{secIII} by posing
\begin{subequations}\label{tx}\begin{align}
	c t &= a(\eta_0) \,(\eta - \eta_0) \, ,\\ z^i &= a(\eta_0) \,(x^i -
	x^i_0) \, ,
\end{align}\end{subequations}
near an event occuring at cosmological time $\eta_0$ and at the galaxy's
center $x^i_0$. In the local coordinate system, the metric developed at the
lowest NR order reads
\begin{subequations}\label{metricNR}\begin{align}
	g_{00} & = -1 + \frac{2 U}{c^2} + \mathcal{O} \left( c^{-4} \right) , \\
	g_{0i} & = \mathcal{O} \left( c^{-3} \right) , \\
	g_{ij} & = \left( 1 + \frac{2 U}{c^2} \right) \delta_{ij} + \mathcal{O} \left( c^{-4} \right) ,
\end{align}\end{subequations}
where $U \ll c^2$ is a Newtonian-like potential. For the motion of massive
(non-relativistic) particules we need only to include the contribution of $U$
in the 00 metric coefficient. Thanks to the standard general relativistic coupling
to gravity in the $ij$ metric coefficient, the motion of photons
agrees with the general relativistic prediction with Newtonian-like potential $U$.

In the NR limit, the equation of motion \eqref{Kdot_Xi} is readily seen to
reduce to
\begin{equation}\label{motion_NR}
	\frac{\ud v^i}{\ud t} - g^i = - \hat{\Pi}_\perp^i \dcalW +
	\mathcal{O} \left( c^{-2} \right) ,
\end{equation}
where $a^i \equiv \ud v^i / \ud t = \left( \partial_t + v^j \partial_j \right) v^i$
is the standard Newtonian acceleration of a fluid in the Eulerian picture,
$v^i$ being the coordinate three-velocity, and $g^i = \partial_i U$ the
non-relativistic local gravitational field. Note that $g^i$ is generated by
both the ordinary baryonic matter and the dipolar dark matter. Similarly, the
equation of evolution \eqref{Omegadot_Xi} for the dipole moment reads in the
NR limit [using also \eqref{motion_NR}]
\begin{equation}\label{evolution_NR}
	\frac{\ud^2 \xi_\perp^i}{\ud t^2} - \hat{\Pi}_\perp^i \dcalW \! =
	\frac{1}{\sigma} \partial_i \left( \calW - \Pi_\perp \dcalW \right) +
	\xi_\perp^j \partial_j g^i + \mathcal{O} \left( c^{-2} \right) ,
\end{equation}
where we explicitly have $ \ud^2 \xi_\perp^i / \ud t^2 = \left( \partial_t^2 +
a^j \partial_j + 2 v^j \partial_{jt}^2 + v^j v^k \partial_{jk}^2 \right)
\xi_\perp^i$. Notice the second term in the RHS which is a tidal term coming
from the Riemann curvature coupling in \eqref{Omegadot_Xi}. Finally, the
equation \eqref{continuity} reduces to the classical continuity equation
\begin{equation}\label{continuity_NR}
	\partial_t \sigma + \partial_i \left( \sigma v^i \right) = \mathcal{O}
	\left( c^{-2} \right) .
\end{equation}

Next, we need to be cautious about the relativistic order of magnitude of the
potential function $\calW$ appearing in the Lagrangian \eqref{L}. It is clear
that $\calW$ has the dimension either of a mass density or an energy density,
depending of where we would reinstall the factors $c$ in \eqref{L}. We shall
from now on assume that $\calW$ is an \textit{energy} density, and has a
finite non-zero limit when $c \rightarrow +\infty$. This will be justified
when we show in \eqref{W23} below that the coefficients $\calW_2$, $\calW_3$,
\dots~in the expansion of $\calW$ considered as an energy density, can be
expressed solely in terms of $G$ and the MOND acceleration $a_0$ (without any
$c$'s). Therefore, our assumption means that we are viewing $a_0$ as a new
fundamental acceleration scale \textit{a priori} independent from $c$. With
such hypothesis, if we reintroduce the factors of $c$ in the expression of the
density $r$ considered as a \textit{mass} density and given by \eqref{r}, we
get $r = \rho + (\calW - \Pi_\perp \dcalW) / c^2$, where $\rho$ is given by
\eqref{gen_rho}. Thus, the term $(\calW - \Pi_\perp \dcalW) / c^2$ becomes
negligible in the formal limit $c \rightarrow +\infty$, and we have $r = \rho
+ \mathcal{O} ( c^{-2})$. In particular, we observe that the term $\calW_0$,
which is linked to the cosmological constant by (restoring the $c$'s and $G$)
\begin{equation}\label{W0Lambda}
	 \calW_0 = \frac{\Lambda c^4}{8 \pi G}\, ,
\end{equation}
does not enter the expression of the dipolar fluid density $r$, and therefore
has no influence on the local dynamics of the dipolar dark matter in the NR
limit. Our assumption that $\calW$ has a finite non-zero limit when $c
\rightarrow +\infty$ means that the cosmological constant $\Lambda$ should
scale with $c^{-4}$, which will be justified later when we show that
$\Lambda\sim a_0^2/c^4$.

Thus, in the NR limit we need to consider only the mass density of the dipolar
dark matter given by $\rho$. Now, from \eqref{gen_rho} we have $\rho = \sigma
- \nabla_\lambda \Pi_\perp^\lambda$ which becomes when $c\rightarrow +\infty$
\begin{equation}\label{rho_NR}
	\rho = \sigma -\partial_i \Pi_\perp^i + \mathcal{O} \left( c^{-2}
	\right) .
\end{equation}
At that order the dipolar term involves only an ordinary partial space
derivative. Finally, we get the Poisson equation in the standard way as the NR
limit of the $00$ and $ii$ components of the Einstein equations, and find
\begin{equation}\label{Poisson1}
	\Delta U = -4 \pi G \,\bigl( \rho_\text{b} + \sigma -\partial_i
	\Pi_\perp^i\bigr) + \mathcal{O} \left( c^{-2} \right) ,
\end{equation}
where $\rho_\text{b}$ is the Newtonian mass density of baryonic matter. This
equation can be written in the alternative form
\begin{equation}\label{Poisson2}
	\partial_i \left( g^i - 4 \pi G \, \Pi_\perp^i \right) = -4 \pi G \,
	\bigl( \rho_\text{b} + \sigma\bigr) + \mathcal{O} \left( c^{-2}
	\right) .
\end{equation}
To summarize, the equations governing the dynamics of the dipolar dark matter 
and the gravitational field in the NR limit are: the equation of motion
\eqref{motion_NR}, the evolution equation \eqref{evolution_NR}, the continuity
equation \eqref{continuity_NR} and the Poisson equation \eqref{Poisson2}. On
the other hand, baryons and photons obey the geodesic equation, which means
$\ud v_\text{b}^i/\ud t = \partial_i U + \mathcal{O} ( c^{-2})$ for baryons,
and the standard GR formula for light deflection in a potential $U$ for
photons, where $U$ is generated by \eqref{Poisson1}.

\subsection{The weak clustering hypothesis}\label{secIVB}

We have shown in section \ref{secIII} that at linear perturbation order, in a
cosmological context, the dynamics of dipolar dark matter cannot be
distinguished from that of baryonic matter or standard dark matter. We now
argue that the motion of dipolar dark matter being non-geodesic, its
\textit{non-linear} dynamics should be different. Our main motivation for the
argument is the existence of an \textit{exact} solution of the equations
governing the dynamics of the dipolar dark matter in the NR limit. Indeed, we
show in appendix \ref{appA} that, in the simple case where the baryonic matter
is modeled by a spherically symmetric mass distribution, there is a solution
to the equations for which the dipole moments are \textit{in equilibrium}
($\xi_\perp = \mathrm{const}$), and \textit{at rest} ($v^i=0$), with the
internal force $\mathcal{F}^i$ exactly balancing the gravitational field
$g^i$. In such a solution, the dipolar medium is uniformly distributed or more
generally spherically symetrically distributed, and the polarization
$\Pi_\perp^i$ is aligned with the gravitational field $g^i$; the dipolar fluid
is thus polarized. Furthermore, we show in this appendix that the latter
solution is stable against dynamical perturbations.

\begin{figure}
	\includegraphics[width=16cm]{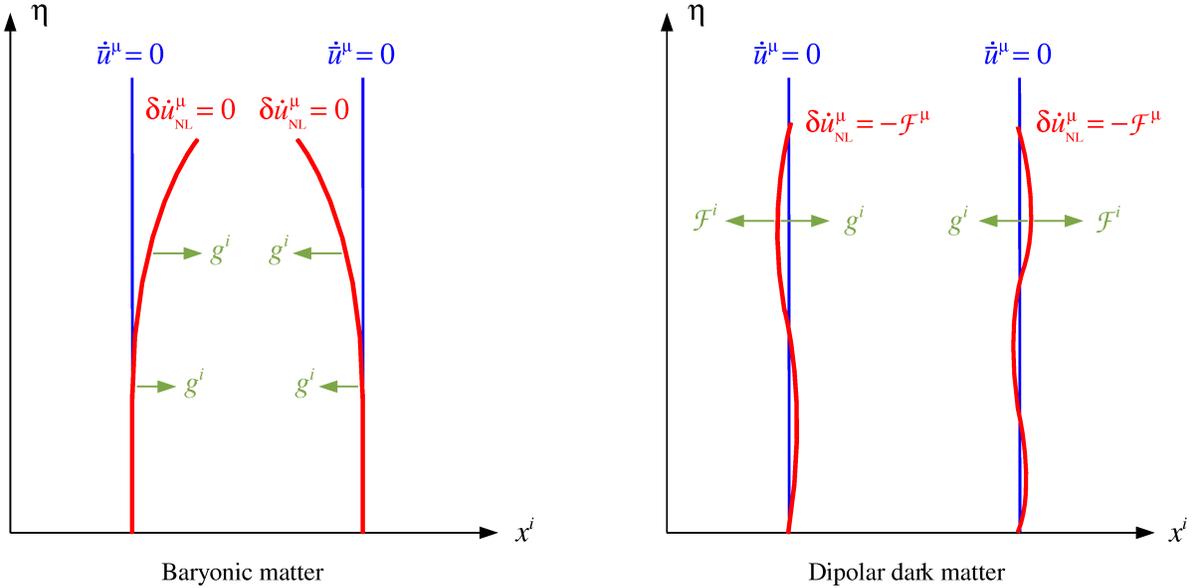}
	\caption{\footnotesize Schematic view of two worldlines of baryonic
	matter and dipolar dark matter. The baryonic matter follows a geodesic
	motion, $\dot{u}^\mu = 0$, and therefore collapses in the regions of
	overdensity. Obeying the non-geodesic equation of motion $\dot{u}^\mu
	= - \mathcal{F}^\mu$, the dipolar dark matter is expected to have a
	different behavior in the \textit{non-linear} (NL) regime. Namely, the
	internal force $\mathcal{F}^i$ can balance the gravitational field
	$g^i$ created by an overdensity, in order to keep the rest mass
	density of dipolar dark matter close to its mean cosmological value,
	$\sigma \sim \overline{\sigma}$, or at least far smaller than the
	baryonic one.}
	\label{motion}
\end{figure}

From that solution, we expect that the dipolar medium will not cluster much
during the cosmological evolution because the internal force may balance part
of the local gravitational field generated by an overdensity (see
Fig.~\ref{motion} for a picturial view of the argument). From this we infer
that the dark matter density contrast in a typical galaxy at low redshift
should be small, at least smaller than in the standard $\Lambda$-CDM
scenario. Such a galaxy would therefore be essentially baryonic, with a
typical mass density of the dipolar dark matter $\sigma$ rather small compared
to the baryonic one, and perhaps around its mean cosmological value
$\overline{\sigma}$. Thus, the crucial hypothesis we are making (based on the
solution in appendix \ref{appA}) is that
\begin{equation}\label{WC1}
	\sigma \ll \rho_\text{b}\, ,
\end{equation}
or perhaps that $\sigma$ stays essentially at a value of the order of its mean
cosmological value,
\begin{equation}\label{WC2}
	\sigma \sim \overline{\sigma} \ll \rho_\text{b}\, .
\end{equation}
Note that for standard CDM (or baryonic matter), the density contrast between
the value of $\rho_\text{cdm}$ (or $\rho_\text{b}$) in a galaxy and the mean
cosmological one $\overline{\rho}_\text{cdm}$ (or $\overline{\rho}_\text{b}$)
is typically of order $10^5$. This means that even if dipolar dark matter
clustered enough so that for instance $\sigma \sim 10^3~\overline{\sigma}$ in
a galaxy at low redshift, it would still satisfy the condition
\eqref{WC1}. 

Note also that with this hypothesis, the non-linear growth of structures in
our model will not be triggered by the rest mass $\sigma$ of dipolar dark
matter (since it does not cluster much), but by the \textit{internal} energy
$\rho_\text{int}$ of the dipolar medium, which is such that
$\rho=\sigma+\rho_\text{int}$ and is explicitly given by $\rho_\text{int} = -
\nabla_\lambda \Pi_\perp^\lambda$ [recall \eqref{gen_rho}]. We have seen that,
at first cosmological perturbation order, the density contrast associated with
$\rho_\text{int}$ reduces to $-D^i\lambda_i$, and obeys the standard evolution
equation \eqref{Dilambdai}. We expect that at non-linear order it will take
over the dominant role as compared to the rest mass density contrast
$\varepsilon$ in the formation of structures. On the other hand, in the NR
limit $\rho_\text{int}$ reduces to $-\partial_i \Pi_\perp^i$ [see
\eqref{rho_NR}] and, as we shall see in the following section, will be at the
origin of the MOND effect.

We shall refer to the condition \eqref{WC1} [or even to the stronger condition
\eqref{WC2}] as the hypothesis of \textit{weak clustering} of the dipolar dark
matter fluid. Obviously, the validity of this hypothesis cannot be addressed
with the formalism of first-order cosmological perturbations in section
\ref{secIII}, because it is a consequence of the non-linear cosmological
evolution. The hypothesis of weak clustering of dipolar dark matter should be
validated through numerical N-body simulations.

Let us thus assume that the dipolar dark matter has not clustered very much,
and even that $\sigma$ might stay more or less at the cosmological mean value
$\overline{\sigma}$ (such that $\Omega_\text{dm}\simeq 0.23$). Because of its
size and typical time-scale of evolution, a galaxy is almost unaffected by the
cosmological expansion of the Universe. Therefore, the cosmological mass
density $\overline{\sigma}$ of the dipolar dark matter is not only
homogeneous, but also almost constant in this galaxy. Thus, the continuity
equation \eqref{continuity_NR} reduces to $\partial_i \left( \overline{\sigma}
v^i \right) \simeq 0$. The most simple solution obviously corresponds to a
static fluid verifying
\begin{equation}\label{rest}
	v^i \simeq 0\, .
\end{equation}
It is therefore natural to consider that the dipolar dark matter is almost at
rest with respect to some averaged cosmological matter distribution. This is
supported by the exact solution found in appendix \ref{appA}, which indicates
that the dipolar dark matter in presence of an ordinary mass does indeed
behave essentially like a static medium. Because of the internal force, the
motion is not geodesic, and the force acts like a ``rocket'' to compensate the
gravitational field and to keep the dipolar particle at rest with respect to
ordinary matter (see Fig.~\ref{motion}).

\subsection{Link with the phenomenology of MOND}\label{secIVC}

Let us now show that under the weak clustering hypothesis, the equations
\eqref{motion_NR}--\eqref{continuity_NR} and \eqref{Poisson2} naturally
reproduce the phenomenology of MOND. First of all, if \eqref{rest} holds,
equation \eqref{motion_NR} tells that the polarization $\Pi_\perp^i$ should be
aligned with the local gravitational field $g^i$, namely\footnote{From now on,
we no longer indicate the neglected remainder terms $\mathcal{O} (
c^{-2})$. Furthermore we assume for the discussion that \eqref{rest} is
exactly verified, i.e. $v^i=0$.}
\begin{equation}\label{gi_Wprime}
        g^i = \hat{\Pi}_\perp^i \dcalW .
\end{equation}
This proportionality relation will be the crucial ingredient for recovering
MOND.

We must now further specify the ``fundamental'' potential $\calW$ entering the
original action \eqref{L}. In section \ref{secIII}, we considered the dipolar
fluid at early cosmological times, where the polarization field was
perturbative. We shall now consider it at late cosmological times (around the
value $\eta_0$) but still in a regime where the polarization field is
weak. This will correspond to the outer zone of a galaxy at low redshift,
where the local gravitational field generated by the galaxy is weak. We
therefore assume that the potential $\calW$ can still be expanded in powers of
$\Pi_\perp$ and we keep only a few terms in the expansion. Next, we introduce
a fundamental acceleration scale $a_0$ to be later identified with the MOND
constant acceleration whose commonly accepted value is $a_0 \simeq 1.2 \times
10^{-10}~\text{m} / \text{s}^2$~\cite{SaMc.02}. Associated with $a_0$ we can
define a fundamental surface density scale
\begin{equation}\label{Sigma}
\Sigma \equiv \frac{a_0}{2 \pi G} \, ,
\end{equation}
whose numerical value is $\Sigma \simeq 0.3~\text{kg}/\text{m}^2 \simeq
130~\text{M}_\odot/\text{pc}^2$. The numerical value of $\Sigma$ is close to
the observed upper limit of the surface brightness of spiral galaxies --- the
so-called Freeman's law which is seen as an empirical evidence for
MOND~\cite{SaMc.02}. We now assert that the expansion of $\calW$ when
$\Pi_\perp\rightarrow 0$ is physically valid when the condition $\Pi_\perp \ll
\Sigma$ is satisfied. As will become obvious, this condition can equivalently
be written $g \ll a_0$, where $g=\vert g^i\vert$ is the norm of the local
gravitational field of the galaxy, and this will correspond to the deep MOND
regime (see Fig.~\ref{mond}). With respect to the expansion \eqref{Wexp}
already used in cosmology, we shall be able to add an extra term. We now write
this expansion, for $\Pi_\perp \ll \Sigma$, as
\begin{equation}\label{Wexp_NR}
	\calW(\Pi_\perp) = \calW_0 + \frac{1}{2} \calW_2 \, \Pi_\perp^2 +
	\frac{1}{6} \calW_3 \, \Pi_\perp^3 + \mathcal{O}
	\Bigl[\left(\hbox{$\frac{\Pi_\perp}{\Sigma}$}\right)^4\Bigr] \, .
\end{equation}
The first term $\calW_0$ is related to the cosmological constant $\Lambda$
through \eqref{W0Lambda}. We now show that the next two coefficients $\calW_2$
and $\calW_3$ are uniquely determined if we want to recover the phenomenology
of MOND. Indeed, by inserting \eqref{Wexp_NR} into the relation
\eqref{gi_Wprime} we obtain
\begin{equation}\label{g_Pi}
	g^i = \Pi_\perp^i \left\{ \calW_2 + \frac{1}{2} \calW_3 \, \Pi_\perp +
	\mathcal{O}
	\Bigl[\left(\hbox{$\frac{\Pi_\perp}{\Sigma}$}\right)^2\Bigr]\right\} ,
\end{equation}
which can be inverted to yield the polarization as an expansion in powers of
(the norm of) the gravitational field. Anticipating that $\calW_2
\, \Sigma \sim a_0$, this expansion will be valid whenever $g \ll a_0$. We
obtain
\begin{equation}\label{Pi_g}
	\Pi_\perp^i = \frac{g^i}{\calW_2} \left\{ 1 - \frac{1}{2}
	\frac{\calW_3}{\calW^2_2} \, g + \mathcal{O}
	\Bigl[\left(\hbox{$\frac{g}{a_0}$}\right)^2\Bigr]\right\} .
\end{equation}
Next, following the conventions of~\cite{Bl1.07,Bl2.07}, we introduce the
coefficient of ``\textit{gravitational susceptibility}'' $\chi$ of the dipolar
medium through
\begin{equation}\label{chidef}
	\Pi_\perp^i = - \frac{\chi}{4 \pi G} \, g^i \, .
\end{equation}
Inserting that definition\footnote{Note that this definition is valid in both
MOND and Newtonian regimes whenever the polarization is aligned with the
gravitational field.} into the LHS of the Poisson equation \eqref{Poisson2},
we find
\begin{equation}\label{Poisson_MOND0}
	\partial_i \left[ (1+\chi) \, g^i \right] = -4 \pi G \, \left(
	\rho_\text{b} + \sigma \right) .
\end{equation}
Finally, invoking our hypothesis of weak clustering \eqref{WC1}, or
\eqref{WC2} in the more extreme variant, we can neglect the mass density
$\sigma$ of the dipole moments with respect to the baryonic one,
so we obtain the MOND equation which is generated solely by the
distribution of baryonic matter as~\cite{BeMi.84}
\begin{equation}\label{Poisson_MOND}
	\partial_i \left( \mu \, g^i \right) = -4 \pi G \, \rho_\text{b} \, .
\end{equation}
The MOND function $\mu$ is related to the susceptibility coefficient by
$\mu=1+\chi$ and can actually be interpreted as the ``digravitational''
coefficient of the dipolar medium~\cite{Bl1.07}. Again, let us stress that in
this model we do have some distribution of dark matter $\sigma$ in an ordinary
sense, but we expect its contribution to become negligible in galactic halos
at low redshifts (after cosmological evolution), so that the MOND fit of
rotation curves of galaxies is unaffected by this ``monopolar'' dark
matter.\footnote{However, at the larger scale of clusters of galaxies the monopolar
part of the dipolar medium $\sigma$ may play a role to explain the missing dark
matter in MOND estimates of the dynamical mass \cite{SaMc.02,An.al.06}. Note that in the
present model, the motion of photons, needed to interpret weak-lensing experiments, is
given by the standard general relativistic prediction; see \eqref{metricNR} with potential
$U$ solution of the MOND equation \eqref{Poisson_MOND}.}
The MOND effect is due to the dipolar part of the dark matter
given by the internal energy $\rho_\text{int} = - \partial_i \Pi^i_\perp$.

Now, from astronomical observations we know that the gravitational
susceptibility $\chi$ in the deep MOND regime $g \ll a_0$ should behave like
\begin{equation}\label{chiexp}
	\chi = -1 + \frac{g}{a_0} +
	\mathcal{O}\left(\left[\hbox{$\frac{g}{a_0}$}\right]^2\right) .
\end{equation}
The fact that $\chi$ should be \textit{negative} was interpreted in
the quasi-Newtonian model~\cite{Bl1.07} as an evidence for gravitational
polarization --- the gravitational analogue of the electric polarization in
dielectric media. By inserting \eqref{chiexp} into \eqref{chidef}, and
comparing with the prediction of our model as given by \eqref{Pi_g}, we
uniquely fix the unknown coefficients therein as
\begin{subequations}\label{W23}\begin{align}
	 \calW_2 &= 4 \pi G \, ,\\ \calW_3 &= 32 \pi^2 \frac{G^2}{a_0} \, .
\end{align}\end{subequations}
This, together with $\calW_0$ fixed by \eqref{W0Lambda}, determines the
potential function $\calW$ up to third order from astronomical
observations. As we see, the MOND acceleration $a_0$ enters at third order in
the expansion, and therefore does not show up in the linear cosmological
perturbations of section \ref{secIII}. At third order, the potential $\calW$
deviates from a purely harmonic potential, and $a_0$ can be seen as a measure
of its anharmonicity.

To express $\calW$ in the best way, we prefer using the surface density scale
$\Sigma = a_0/2\pi G$ rather than the acceleration scale $a_0$. To do so, we
must introduce a purely numerical dimensionless coefficient $\alpha$ to
express the cosmological constant $\Lambda$ (which is positive and has the
dimension of an inverse length squared) in units of $a_0^2/c^4$, and we pose
\begin{equation}\label{Lambda_a0}
	\Lambda = 3 \alpha^2 \left( \frac{2\pi a_0}{c^2} \right)^2 .
\end{equation}
The definition of $\alpha$ is such that $a_\Lambda = \alpha\,a_0$ represents
the natural acceleration scale associated with the cosmological constant, and
is already given by \eqref{aLambda} as $a_\Lambda = \sqrt{\Lambda / 3} \, c^2
/ 2 \pi$. Then, the cosmological term \eqref{W0Lambda} becomes $\calW_0 =
6\pi^3 G \, \Sigma^2 \, \alpha^2$, and we obtain
\begin{equation}\label{Wsimple}
	\calW = 6 \pi G\,\Sigma^2 \left\{ \alpha^2\pi^2 +
	\frac{1}{3}\left(\frac{\Pi_\perp}{\Sigma}\right)^2 + \frac{4}{9}
	\left(\frac{\Pi_\perp}{\Sigma}\right)^3 + \mathcal{O}
	\biggl[\left(\frac{\Pi_\perp}{\Sigma}\right)^4\biggr] \right\} .
\end{equation}
In the present model there is nothing which can give the relation between
$\Lambda$ and $a_0$, hence $\alpha$ is not determined. However, if the dipolar
fluid action \eqref{L} is intended to describe at some macroscopic level a
more fundamental theory (presumably a QFT), we expect that the potential
$\calW$ should depend only on certain more or less fundamental constants, and
some dimensionless variables built from ``fundamental fields''. Introducing
the dimensionless quantity $x \equiv \Pi_\perp / \Sigma$, we can rewrite
\eqref{Wsimple} as $\calW = 6\pi G \, \Sigma^2 \, w(x)$, where
\begin{equation}\label{wfonstion}
	w(x)=\alpha^2\pi^2 + \frac{1}{3}x^2 + \frac{4}{9}x^3 +
	\mathcal{O}(x^4)
\end{equation}
represents some ``universal'' function coming from some fundamental albeit
unknown physics. Therefore, we expect that the numerical coefficients in the
expansion of $w(x)$ should be of the order of one or, say, 10. In particular,
it is natural to expect that $\alpha$ should be of the order of one (to within
a factor $10$ say), and we deduce from \eqref{Lambda_a0} that the magnitude of
$\Lambda$ should scale approximately with the square of the MOND acceleration,
namely $\Lambda \sim a_0^2/c^4$.

The numerical coincidence between the measured values of $\Lambda$ and $a_0$
is well-known~\cite{Mi.02}. The observed value of the cosmological constant is
around $\Lambda \simeq 0.12~\text{Gpc}^{-2}$~\cite{PeUz} which, together with
$a_0 \simeq 1.2 \times 10^{-10}~\text{m} / \text{s}^2$, corresponds to a value
for $\alpha$ which is very close to one: $\alpha \simeq 0.8$. Thus $a_0$ is
very close to the scale $a_\Lambda$ associated with the cosmological constant,
which is related to the Gibbons-Hawking temperature $T_\text{GH} = \hbar
a_\Lambda/k c$ derived from semi-classical theory on de Sitter
space-time~\cite{GiHa.77}. From the previous discussion, we see that the
``cosmic'' coincidence between $\Lambda$ and $a_0$ has a natural explanation
if dark matter is made of a medium of dipole moments.

\subsection{The Newtonian regime}\label{secIVD}

For the moment, we looked at the explicit expression of the potential function
$\calW$ in the MOND regime $g \ll a_0$. We would also like to get some
information about this function in the Newtonian regime $g \gg a_0$.
To do so, we first derive the general expression of the
gravitational susceptibitity coefficient $\chi$. Here we assume that the MOND
function $\mu = 1 + \chi$ is always less than $1$. This implies $\chi < 0$ and
thus using \eqref{gi_Wprime} and \eqref{chidef} we must have $\dcalW >0$
(where we recall that $\dcalW \equiv \ud \calW / \ud \Pi_\perp$). Taking the
norm of \eqref{gi_Wprime} we get $g = \dcalW \! (\Pi_\perp)$. Next, we
introduce the function $\Theta(g)$ which is the inverse of $\dcalW \!
(\Pi_\perp)$, i.e. satisfies
\begin{equation}\label{Theta}
	g = \dcalW \! (\Pi_\perp) \quad \Longleftrightarrow \quad \Pi_\perp =
	\Theta(g) \, .
\end{equation}
According to \eqref{chidef}, the susceptibility $\chi$ is then given as the
following fonction of the gravitational field $g$,
\begin{equation}\label{chi_P}
	\chi(g) = -4 \pi G \, \frac{\Theta(g)}{g} \, .
\end{equation}
This is the general relation linking $\chi$ (or equivalently the MOND function
$\mu=1+\chi$) to the potential function $\calW$ in the dipolar action
\eqref{L}. Of course, in the present model $\calW$ is to be considered as more
fundamental than $\chi$ which is a derived quantity.

\begin{figure}
\begin{minipage}[c]{.46\linewidth}
		\includegraphics[width=8.5cm]{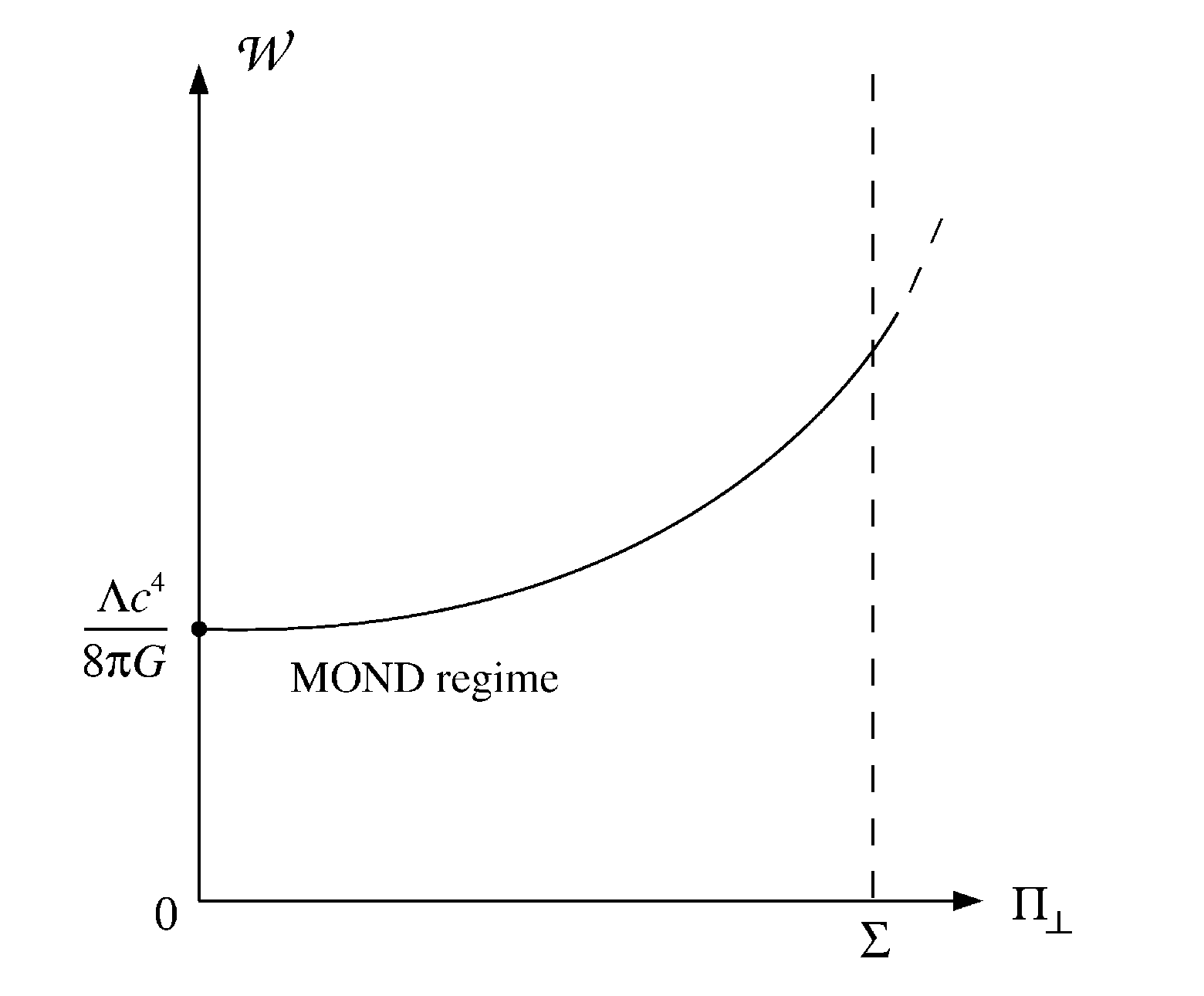}
		\caption{\footnotesize The minimum of the potential function
		$\calW(\Pi_\perp)$, reached when $\Pi_\perp = 0$, is a
		cosmological constant $\Lambda$. Small deviations around the
		minimum, corresponding to $\Pi_\perp \ll \Sigma = a_0/2 \pi
		G$, describe the MOND regime $g \ll a_0$.}
		\label{mond}
	\end{minipage} \hfill
	\begin{minipage}[c]{.46\linewidth}
		\includegraphics[width=8.5cm]{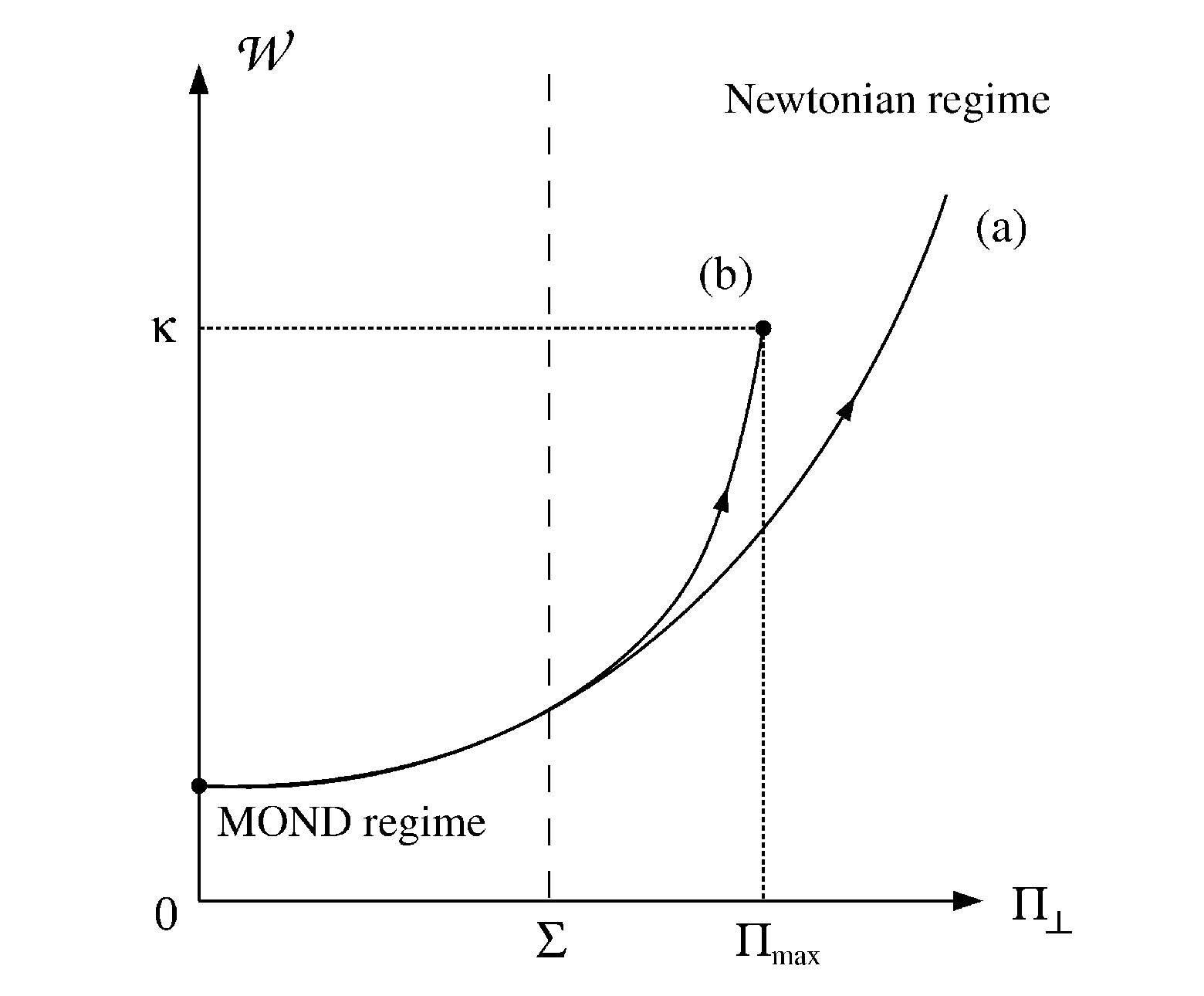}
		\caption{\footnotesize The potential $\calW$ as a function of
		the polarization $\Pi_\perp$ for different asymptotic
		behaviors of the gravitational susceptibility $\chi$ in the
		Newtonian regime $g \gg a_0$. The arrows indicate the
		direction of increasing gravitational field $g$.}
		\label{newtonian} 
\end{minipage} 
\end{figure} 

In the Newtonian regime $g \gg a_0$, the MOND function $\mu$ should tend to
one, so that $\chi$ vanishes in this regime. To discuss more concretely this
condition, we assume that in the formal limit $g\rightarrow +\infty$, the
gravitational susceptibility behaves as $\chi \sim g^{-\gamma}$, with $\gamma$
a strictly positive real number. More precisely, it should behave like $\chi
\sim - \epsilon \left( g/a_0 \right)^{-\gamma}$, where $\epsilon$ is a
strictly positive real number. Beware that even if this power-law
behavior is a simple assumption, nothing garanties that it is verified. Then,
when $g \rightarrow + \infty$, we get from \eqref{Theta} and \eqref{chi_P} that
\begin{subequations}\label{PiWg}\begin{align}
	\Pi_\perp &\sim A \, g^{1-\gamma} \, ,\\ \calW &\sim
	\frac{1-\gamma}{2-\gamma} \, A \, g^{2-\gamma} + \kappa \, ,\label{W_g}
\end{align}\end{subequations}
where $A = \epsilon \, a_0^\gamma / 4 \pi G > 0$ and $\kappa$ is an
integration constant. We have to distinguish several cases, depending on the
value of the exponent $\gamma$:
\begin{itemize}
	\item[(i)] If $0 < \gamma < 1$, then both the polarization $\Pi_\perp$
	and the potential $\calW$ diverge. This would corresponds to the curve
	(a) of Fig.~\ref{newtonian}.
	\item[(ii)] If $\gamma = 1$, the polarization $\Pi_\perp$ tends to a
	maximum ``saturation'' value $\Pi_\text{max} = A$, and the potential
	$\calW$ equals the constant $\kappa$. See curve (b) in
	Fig.~\ref{newtonian}.
	\item[(iii)] If $1 < \gamma < 2$, the polarization goes to zero while
	the potential diverges to $-\infty$ like a power law. This implies
	that $\calW$ cannot be a univalued function of $\Pi_\perp$. Therefore,
	there must exist two branches corresponding to the Newtonian and MOND
	regimes.
	\item[(iv)] If $\gamma = 2$, according to \eqref{W_g} the potential
	diverges to $-\infty$ logarithmically, i.e. $\calW \sim - A \, \ln{g}$, while the
	polarization still vanishes. Same conclusions as in case (iii) apply.
	\item[(v)] Finally, if $ \gamma > 2$, the polarization goes to zero
	while the potential tends to $\kappa$. Same conclusions as in (iii)
	apply.
\end{itemize}
If we believe that the potential $\calW$ represents a fundamental function in
the action, and that our model should strictly speaking be valid in a
Newtonian regime (and not being merely valid in the MOND regime), we should
\textit{a priori} expect that $\calW$ is a univalued function of
$\Pi_\perp$. Then, the susceptibility coefficient should be like $\chi \sim
g^{-\gamma}$ with $0 < \gamma \leqslant 1$ in the Newtonian regime. This would
mean that the MOND function $\mu$ behaves like
\begin{equation}\label{muN}
	\mu \sim 1 - \epsilon \left(
	\frac{a_0}{g} \right)^\gamma ,
\end{equation}
with $0 < \gamma \leqslant 1$. Such rather slow transition of $\mu$ toward the
Newtonian regime is consistent with the recent results of
\cite{FaBi.05} who fitted the rotation curves of the Milky Way and
galaxy NGC~3198, and of~\cite{SaNo.07} who fitted 17 early-type disc galaxies,
and concluded that the Newtonian regime is rather slowly reached. For
instance, the authors of~\cite{FaBi.05,SaNo.07,ZhFa.06} agreed that $\gamma =
1$ yields a better fit to the data than $\gamma = 2$.

The case $\gamma = 1$ (curve (b) in Fig.~\ref{newtonian}) corresponds to an interesting physical situation in
which the dipolar medium \textit{saturates} when $g \rightarrow + \infty$, at
the maximum value $\Pi_\text{max} = A$, or
\begin{equation}
	\Pi_\text{max} = \frac{\epsilon}{2} \, \Sigma \, ,
\end{equation}
where $\Sigma$ is the surface density scale \eqref{Sigma}. In this saturation
case, the gravitational susceptibility coefficient behaves as
\begin{equation}\label{chidom}
	\chi \sim - \epsilon \, \frac{a_0}{g}\, .
\end{equation}

However, let us remind that such a slow transition from MOND toward the
Newtonian regime is \textit{a priori} ruled out by Solar System
observations. Indeed, according to the MOND equation, a planet orbiting the
Sun feels a gravitational field $g$ obeying $(1+\chi)g=g_\text{N}$, where
$g_\text{N}$ is the Newtonian gravitational field. Hence, if $\chi$ scales
like $g^{-1}$ when $g \gg a_0$ like in \eqref{chidom}, the gravitational field
experienced by planets will involve a constant supplementary acceleration
directed toward the Sun (i.e. a ``Pioneer-type'' anomaly) given by
\begin{equation}\label{ggN}
	g \sim g_\text{N} + \epsilon \, a_0 \, .
\end{equation}
Of course it is striking that the order of magnitude of the Pioneer anomaly is
the same as the MOND acceleration $a_0$. Unfortunately, the presence of a
constant acceleration such as in \eqref{ggN} should be detected in the motion
of planets, and this is incompatible with current measurements (see
e.g.~\cite{JaRe1.05,JaRe2.05} for a discussion).

Despite the fact that a slow transition to the Newtonian regime (like for
example the case $\gamma=1$) seems to be favored by observations at the
galactic scale~\cite{FaBi.05,ZhFa.06,SaNo.07}, it does not seem to be viable
when extrapolated up to the scale of the Solar System. In our model, we found
that such a behavior is the result of our belief that the ``fundamental''
function $\calW$ be univalued. In this respect, the validity of the model
should be limited to large scales, from the galactic scale up to cosmological
scales, i.e. in a regime of weak gravity. At smaller scales the description in
terms of a single univalued function $\calW$ should break down. But our model
being an effective one, or even a phenomenological one, the question of
whether the potential $\calW$ is univalued or not remains an open issue.

\section{Summary and conclusion}\label{secV}

In this paper, we proposed a model of dark matter and dark energy based on the
concept of gravitational polarization of a medium of dipole moments. The
dynamics of the dipolar fluid is governed by the Lagrangian \eqref{L} in
standard general relativity, and constitutes a generalization of the previous
model~\cite{Bl2.07}. Namely, this Lagrangian involves a potential function
$\calW$, describing at some effective level a non-gravitational internal force
influencing the dynamics of the dipolar fluid, and which depends on the
polarization field or density of dipole moments $\Pi_\perp = \sigma \xi_\perp$
instead of merely the dipole moment itself $\xi_\perp$ in the
model~\cite{Bl2.07}. This new form of the potential permits recovering in a
most elegant way the phenomenology of MOND in a typical galaxy at low
redshift. In addition, we show that the model naturally contains a
cosmological constant $\Lambda$.

We proved in section \ref{secIII} that whithin the framework of the theory of
first-order cosmological perturbations, the dipolar fluid behaves exactly as
standard dark energy (i.e. a cosmological constant) plus standard dark matter
(i.e. a pressureless perfect fluid). Thus, our model is consistent with the
cosmological observations at large scales. In particular, it leads to the same
predictions as the standard $\Lambda$-CDM model for the CMB
fluctuations. However, at second order in the cosmological perturbations, we
expect that the dipolar dark matter should differ from a perfect fluid because
of the influence of the internal force resulting in a
non-geodesic motion. The model could thus be checked by
working out the second-order cosmological perturbations and comparing with CMB
fluctuations (notably the effects linked with the non-gaussianity).

The dynamics of the dipolar dark matter being different from that of standard
dark matter (at the level of non-linear perturbations), we expect the
``monopolar'' part of the dipolar dark matter not to cluster much during the
cosmological evolution. We call this expectation the hypothesis of ``weak
clustering''. It is supported by an exact solution worked out in appendix
\ref{appA} for the dynamics of dipolar dark matter in the non-relativistic
limit and in spherical symmetry. In this solution, the internal force balances
the local gravitational field produced by a spherical mass, so that the dark
matter remains at rest with respect to the central mass. The weak clustering
hypothesis should be checked \textit{via} N-body numerical simulations. Under
that hypothesis, we show that the Poisson equation for the gravitational field
generated by the baryonic and dipolar dark matter reduces to the MOND equation
in the regime of weak gravitational fields $g \ll a_0$. Our model of dipolar
dark matter therefore naturally explains all the successes of the MOND
phenomenology.

To achieve this result (in section \ref{secIV}) we have to adjust the
fundamental potential $\calW$ in the action. We find that it should be given
by an anharmonic potential, the minimum of which, reached when $\Pi_\perp=0$,
being directly related to the cosmological constant $\Lambda$. It is tempting
to interpret $\Lambda$ as a ``vacuum polarization'' of some hypothetical
quantum field, when the ``classical'' part of the polarization $\Pi_\perp
\rightarrow 0$. The expansion around that minimum is fine-tuned in order to
recover MOND. In particular, we show that the MOND acceleration $a_0$
parametrizes the coefficient of the third-order deviation of $\calW$ from the
minimum. Although fine-tuned to fit with observations, this potential function
$\calW$ offers a nice unification between the dark energy in the form of
$\Lambda$ and the dark matter in the form of MOND (see Fig. \ref{mond}). A
consequence of such unification is that the cosmological constant should scale
with the MOND acceleration according to $\Lambda \sim a_0^2 / c^4$. This
scaling relation is in good agreement with observations and has a very natural
explanation in our model.

To conclude, we proposed to modify the matter sector rather than the gravity
sector as in modified gravity theories \cite{Be.04,Sa.05,Zh.07,Ha.al.08}. Namely,
we investigated a model of dark matter, but of such an exotic form that it naturally explains the
phenomenology of MOND at galactic scales. Furthermore, that form of dark matter
has a simple physical interpretation in terms of the well-known mechanism of
polarization by an exterior field. More work is necessary to test the
model, either by studying second-order perturbations in cosmology, or by
computing numerically the non-linear growth of perturbations and comparing
with large-scale structures.

\acknowledgments 

It is a pleasure to thank Alain Riazuelo and Jean-Philippe Uzan for
interesting discussions at an early stage of this work.

\appendix

\section{Dark matter in a central gravitational field}\label{appA}

We investigate the dynamics of the dipolar dark matter fluid in presence of a
spherically symmetric mass distribution of ordinary baryonic matter in the NR
limit $c \rightarrow +\infty$. The equations to solve are the equation of
motion \eqref{motion_NR}, the equation of evolution \eqref{evolution_NR}, the
continuity equation \eqref{continuity_NR} and the Poisson equation for the
gravitational field \eqref{Poisson2}. Let us rewrite those equations here for
convenience:\footnote{In this appendix, we adopt 3-dimensional notations with
boldface vectors, e.g. $\bm{\mathcal{F}}=(\mathcal{F}^i)$. We also remove the
subscript $\perp$ from the variables $\bm{\xi}_\perp$ and $\bm{\Pi}_\perp$ for
notational simplicity. The derivatives of the potential $\calW$ with respect
to its argument $\Pi$ will be denoted with a prime,
e.g. $\calW'\equiv\calW_{{}_{\Pi}}\equiv\ud\calW/\ud\Pi$.}
\begin{subequations}\label{eqs_stat}
	\begin{align}
		\frac{\ud \bm{v}}{\ud t} &= \bm{g} - \bm{\mathcal{F}} \, ,
		\label{motion_stat} \\ \partial_t \sigma &= - \bm{\nabla}
		\cdot \left( \sigma \bm{v} \right) , \label{cont_stat} \\
		\bm{\nabla} \cdot \bm{g} &= - 4 \pi G \, \left( \sigma +
		\rho_\text{b} - \bm{\nabla} \cdot \bm{\Pi} \right) ,
		\label{poiss_stat} \\ \frac{\ud^2 \bm{\xi}}{\ud t^2} 
                &= \bm{\mathcal{F}} + \frac{1}{\sigma} \bm{\nabla} \left(
		\calW - \Pi \, \calW' \right) + \left( \bm{\xi} \cdot
		\bm{\nabla} \right) \bm{g} \, ,
		\label{evol_stat}
	\end{align}
\end{subequations}
where the internal force reads $\bm{\mathcal{F}} = \hat{\bm{\Pi}} \, \calW'$,
with $\hat{\bm{\Pi}}\equiv\bm{\Pi}/\Pi$.

Our aim is to solve the equations \eqref{eqs_stat} in the special case where
the baryonic matter is modeled by a time-independent spherically symmetric
distribution of mass $\rho_\text{b}(r)$, say with compact support. Let us show
that there is a simple solution to such a set of equations, in the case where
\begin{subequations}\label{sol_stat}
	\begin{align}
		\bm{v}_0 &= \bm{0} \, , \label{vstat}\\ \sigma_0 &= \sigma_0(r) \, ,
	\end{align}
\end{subequations}
which corresponds to a static fluid whose mass distribution is
time-independent and spherically symmetric. We denote such particular solution
with a lower index $0$. From \eqref{sol_stat} we observe that the continuity
equation \eqref{cont_stat} is immediately satisfied. In such a solution,
according to \eqref{motion_stat} the internal force balances exactly the
gravitational field, i.e. $\bm{\mathcal{F}}_0=\bm{g}_0$ (this is somewhat
similar to the case of a non-rotating star in hydrostatic equilibrium, where
the pressure gradient plays the role of the internal force). We deduce that
the polarization field $\bm{\Pi}_0 = \sigma_0 \, \bm{\xi}_0$ is aligned with
the gravitational field $\bm{g}_0$. Hence, from equation \eqref{poiss_stat}
both $\bm{\Pi}_0$ and $\bm{g}_0$ are radial. We shall pose $\bm{g}_0 = -
g_0(r,t) \, \bm{e}_r$ and $\bm{\Pi}_0 = - \Pi_0(r,t) \, \bm{e}_r$, where in
our notation $g_0 > 0$ and $\Pi_0 > 0$. 

Furthermore, let us show that in addition the polarization field is practically in
``equilibrium'', i.e. $\Pi_0$ is independent on time $t$, and so is $g_0$. We
replace $\bm{g}_0$ by the explicit expression of the internal force
$\bm{\mathcal{F}}_0=\hat{\bm{\Pi}}_0 \calW'_0$ into the evolution equation
\eqref{evol_stat}, use \eqref{vstat} and get
\begin{equation}\label{xipp}
	\partial_t^2\bm{\Pi}_0 - \sigma_0 \calW'_0 \, \hat{\bm{\Pi}}_0 =
	\bm{\nabla} \left( \calW_0 - \Pi_0 \, \calW_0' \right) + \left(
	\bm{\Pi}_0 \cdot \bm{\nabla} \right) (\hat{\bm{\Pi}}_0 \, \calW'_0) \,
	.
\end{equation}
Here $\hat{\bm{\Pi}}_0=\bm{\Pi}_0/\Pi_0=-\bm{e}_r$, and we introduced the
shorthand notation $\calW_0' \equiv \calW'(\Pi_0)$. Now, it turns out that the
RHS of this equation vanishes in the special case where the polarization field
is radial, hence we get
\begin{equation}\label{xipp2}
	\partial_t^2 \Pi_0 = \sigma_0 \calW'_0 \, .
\end{equation}
In order to determine the time evolution of $\Pi_0$, an explicit expression
for the potential $\calW$ is in principle required. However, we saw in section
\ref{secIVC} that the potential $\calW$ only depends on the polarization $\Pi$
and the constants $a_0$ and $G$. The only time-scale one can build with $a_0$,
$G$ and $\sigma_0$ is the dipolar dark matter self-gravitating time-scale
$\tau_\text{g} = (\pi/G \sigma_0)^{1/2}$, or equivalently, in terms of
frequency, $\omega_\text{g}^2 = 4 \pi G \sigma_0$. Therefore, the polarization
$\Pi_0$ can only evolve on this time-scale. For instance, in the MOND regime
$g\ll a_0$, we have at leading order $\calW'_0=4\pi G \,\Pi_0$, hence
\eqref{xipp2} reduces to
\begin{equation}\label{dttPi0}
	\partial_t^2 \Pi_0 = \omega_\text{g}^2 \, \Pi_0 \, .
\end{equation}
The most general solution of this equation is a linear combinaison of
hyperbolic $\cosh{\omega_\text{g} t}$ and $\sinh{\omega_\text{g} t}$. For a
``monopolar'' dark matter mass density $\sigma_0$ of, say, the mean cosmological
value $\overline{\sigma} \simeq 10^{-26}~\text{kg}/\text{m}^3$ [in agreement with
our weak clustering hypothesis \eqref{WC2}], the typical time-scale of
evolution of $\Pi_0$ will be larger than $6 \times 10^{10}~\text{years}$. This is
large enough to neglect any time variation of $\Pi_0$ with respect to a typical
orbital time-scale in a galaxy. Our solution is therefore given by
\begin{equation}\label{Pi0}
	\bm{\Pi}_0 = - \Pi_0(r) \, \bm{e}_r\, ,
\end{equation}
together with \eqref{sol_stat}. The dipole moments are at rest and in
equilibrium. The explicit function $\Pi_0(r)$ is determined from the radial
gravitational field $g_0(r)$ as\footnote{Note that if in this solution the
polarization field $\Pi_0(r)=\sigma_0(r)\xi_0(r)$ is determined, the
density $\sigma_0(r)$ and dipole moment $\xi_0(r)$ are not specified
separately. For instance, the density could be at the uniform cosmological
value $\overline{\sigma}$ so that $\xi_0(r)=\Pi_0(r)/\overline{\sigma}$. This
degeneracy of $\sigma_0(r)$ is an artifact of our assumptions of spherical
symmetry and staticity.}
\begin{equation}\label{Pi0r}
	\Pi_0(r) = \Theta\left(g_0(r)\right)\, ,
\end{equation}
where $\Theta(g_0)$ denotes the inverse inverse function of $\calW'(\Pi_0)$
following the notation \eqref{Theta}. The gravitational field $g_0(r)$ is
determined by the Poisson equation \eqref{poiss_stat} as
\begin{equation}\label{g0r}
	g_0 - 4\pi G \,\Pi_0 = \frac{G M_0(r)}{r^2}\, ,
\end{equation}
where $M_0(r) = 4\pi\int_0^r\ud s\,s^2[\rho_\text{b}(s) + \sigma_0(s)]$ is the
mass enclosed within radius $r$.

The existence of this physically simple solution represents a notable progress
compared to the more complicated solution found in the previous
model~\cite{Bl2.07} (see section IV there). Such a solution is quite
interesting for the present model because it indicates that during the
cosmological evolution (at non-linear perturbation order) the dipolar dark
matter may not cluster very much toward regions of overdensity. Most of the
effect will be in the dipole moment vectors which acquire a spatial
distribution. This is our motivation for the ``weak clustering'' assumption
\eqref{WC1}--\eqref{WC2} stating that $\sigma \ll \rho_\text{b}$, which was
used in section \ref{secIVC} to obtain MOND. In the present case, neglecting
$\sigma_0$ with respect to $\rho_\text{b}$ in the RHS of \eqref{g0r}, we
recover the usual MOND equation generated by the baryonic density only. This
being said, such an appealing solution may be physically irrelevant if the
spherically symmetric configuration appears to be unstable with respect to
linear perturbations. This motivates the following study of the stability of
the previous solution.

Consider a general perturbation of the background solution, namely
\begin{subequations}
	\begin{align}
		\sigma &= \sigma_0 + \delta \sigma \, , \\ \bm{v} &=
		\delta\bm{v} \, , \\ \bm{\Pi} &= \bm{\Pi}_0 + \delta\bm{\Pi}
		\, .
	\end{align}
\end{subequations}
We have also $\bm{g} = \bm{g}_0 + \delta\bm{g}$ and $\bm{\mathcal{F}} =
\bm{\mathcal{F}}_0 + \delta\bm{\mathcal{F}}$, where the expression of the
perturbed force in terms of the perturbed polarization explicitly reads
\begin{equation}
	\delta\bm{\mathcal{F}} = \calW_0''\,(\hat{\bm{\Pi}}_0 \cdot
	\delta\bm{\Pi}) \, \hat{\bm{\Pi}}_0 + \calW_0' \left[
	\frac{\delta\bm{\Pi}}{\Pi_0} - \left( \hat{\bm{\Pi}}_0 \cdot
	\frac{\delta\bm{\Pi}}{\Pi_0} \right) \hat{\bm{\Pi}}_0 \right] .
\end{equation}
Assuming a Fourier decomposition for any perturbative quantity $\delta X$, we
write for a given mode of frequency $\omega$ and wave number $\bm{k}$,
\begin{equation}\label{perturb}
	\delta X (\bm{x},t) = \delta X (\bm{k},\omega) \, e^{\ui (\bm{k} \cdot
	\bm{x} - \omega t)} \, .
\end{equation}
We want to find the relation between $\bm{k} \cdot \bm{e}_r$ and $\omega$, the
so-called dispersion relation, which contains all the physical information
about the behavior of the generic perturbation \eqref{perturb}. Introducing
this \textit{ansatz} into \eqref{eqs_stat}, and simplifying the resulting
equations by making use of the background solution, we find
\begin{subequations}
	\begin{align}
		\delta\bm{v} &= \frac{\ui}{\omega} \left( \delta\bm{g} -
		\delta\bm{\mathcal{F}} \right) , \label{motion_pert} \\ \delta
		\sigma &= \frac{1}{\omega} \left( \sigma_0 \,\bm{k} \cdot
		\delta\bm{v} - \ui \,\delta\bm{v} \cdot \bm{\nabla} \sigma_0
		\right) , \label{cont_pert} \\ \delta\bm{g} &= 4 \pi G
		\,\frac{\ui \,\bm{k}}{k^2} \, \left( \delta \sigma - \ui \,\bm{k}
		\cdot \delta\bm{\Pi} \right) .
	\end{align}
\end{subequations}
These algebraic expressions can be combined to express $\delta \sigma$,
$\delta\bm{g}$ and $\delta\bm{v}$ in terms of $\delta\bm{\Pi}$ only. After
some algebra, we get from the evolution equation \eqref{evol_stat} a relation
expressing the perturbed polarization field $\delta\bm{\Pi} = \sigma_0 \,
\delta\bm{\xi} + \delta \sigma \, \bm{\xi}_0$ as
\begin{align}\label{master}
	\omega^2 \, \delta\bm{\Pi} &= \omega^2 \, \frac{\delta
	\sigma}{\sigma_0} \, \bm{\Pi}_0 + \frac{\ui \,\omega}{\sigma_0} \left(
	\delta\bm{v} \cdot \bm{\nabla} \sigma_0 \right) \bm{\Pi}_0 -
	\ui\,\omega \left( \delta\bm{v} \cdot \bm{\nabla} \right) \bm{\Pi}_0 +
	(\hat{\bm{\Pi}}_0 \cdot \delta\bm{\Pi}) \, \bm{\nabla} \left( \Pi_0
	\calW_0'' \right) \nonumber \\ &+ \Pi_0 \calW_0'' \, (\hat{\bm{\Pi}}_0
	\cdot \delta\bm{\Pi}) \, \ui\,\bm{k} - \left( \ui\,\bm{k} \cdot
	\bm{\Pi}_0 \right) \delta\bm{g} - \left( \delta\bm{\Pi} \cdot
	\bm{\nabla} \right) \bm{g}_0 - \sigma_0 \, \delta \bm{\mathcal{F}} \, .
\end{align}
When replacing $\delta \sigma$, $\delta\bm{g}$, $\delta\bm{v}$ and $\delta
\bm{\mathcal{F}}$ into \eqref{master} we obtain a master equation for the
perturbed polarization $\delta\bm{\Pi}$ which is quite complicated. Given the
complexity of the problem, we restrict our analysis to the simplest modes in a
spherically symmetric background, namely those propagating radially. We shall
thus write $\bm{k} = k \, \bm{e}_r$, and study successively the transverse and
longitudinal perturbations.

Firstly, let us consider a transverse perturbation $\delta\bm{\Pi}$, i.e. one
which satisfies $\delta\bm{\Pi} \cdot \bm{e}_r = 0$. Projecting the master
equation \eqref{master} in the direction of $\delta\bm{\Pi}$, we get that
\begin{equation}
	\left[\omega^2 + \calW'_0
	\left(\frac{1}{\xi_0}-\frac{2}{r}\right)\right] \delta \Pi = 0 \, ,
\end{equation}
which simply states that no transverse perturbations propagating radially are
allowed, i.e. $\delta \Pi = 0$. Consider now the case of a longitudinal
perturbation $\delta\bm{\Pi} = - \delta \Pi(r,t) \, \bm{e}_r$, where $\delta
\Pi$ can be positive or negative (with our convention the norm of $\bm{\Pi}$
reads $\Pi=\Pi_0+\delta\Pi$), and represents the arbitrary amplitude of the
applied linear perturbation. After some lengthy calculations, we get the
dispersion relation
\begin{equation}
	k = \ui \, \frac{\partial_r \sigma_0}{\sigma_0} \left( 1 +
	\frac{\omega^2}{\omega_\text{g}^2} \left[ 1 + \frac{\left( 4 \pi G -
	\calW_0'' \right) \partial_r \Pi_0}{\omega^2 + \sigma_0 \calW''_0 +
	\Pi_0 \, \partial_r \calW_0''}\right] \right)^{-1} .
\end{equation}
Notice first that, as the wave number $k$ is purely imaginary, such a
perturbation cannot propagate. Secondly, the stability of the background
solution with respect to this perturbation is related to the sign of
$k/\ui$, so an explicit expression for the potential $\calW$ is required to
conclude. Such an expression is available in the MOND regime $g_0 \ll a_0$
using the expansion \eqref{Wsimple}. Assuming the MOND equation with a
(baryonic) point mass $M$ for simplicity, i.e. equation \eqref{g0r} with
$\rho_\text{b}=M\,\delta(\bm{x})$ and negligible $\sigma_0$, we find that the
dispersion relation can be recast at the leading order in the form
\begin{equation}\label{disp_mond}
	k = \ui \, \frac{\partial_r \sigma_0}{\sigma_0} \,
	\frac{\omega_\text{g}^2 \left( \omega^2 + \omega_\text{g}^2 - 2 \,
	\omega^2_\text{K} \right)}{\omega^4 + 2 \, \omega_\text{g}^2 \,
	\omega^2 + \omega_\text{g}^2 \left( \omega_\text{g}^2 - 2 \,
	\omega_\text{K}^2 \right)} \, ,
\end{equation}
where $\omega_\text{K}^2 = GM/r^3$ denotes the Keplerian frequency. We now
turn to the analysis of the two factors in \eqref{disp_mond}, namely the
$\omega$-dependent and $\sigma_0$-dependent ones.

At a given distance $r$ from the center of the galaxy, the $\omega$-dependent
factor becomes very large in the vicinity of the resonant frequency

\begin{equation}\label{omega_r}
	\omega_\text{R}^2 = \omega_\text{g} \left( \sqrt{2} \, \omega_\text{K}
	- \omega_\text{g} \right).
\end{equation}
But we are restricting our attention to perturbations in the MOND regime where
$g_0 \ll a_0$, which means at distances $r$ from the galactic center that are
far larger than the MOND radius $r_\text{M} \equiv \sqrt{GM/a_0}$, or
equivalently at Keplerian frequencies $\omega_\text{K} \ll \omega_\text{M}$
with $\omega^2_\text{M}=G M/r^3_\text{M}$. For a typical galaxy of mass $M
\sim 10^{11}~\text{M}_\odot$, and a ``monopolar'' dark matter mass density
around the mean cosmological value, i.e. $\sigma_0 \sim \overline{\sigma}
\simeq 10^{-26}~\text{kg}/\text{m}^3$, we find by reporting the constraint
$\omega_\text{K} \ll \omega_\text{M}$ into \eqref{omega_r} the upper-bound
$\omega^2_\text{R} \ll \sqrt{2} \, \omega_\text{g} \, \omega_\text{M}$, which
gives numerically $\omega_\text{R} \ll 10^{-17}~\text{s}^{-1}$. Because
perturbations with a typical time scale $2\pi/\omega \gg 2 \times
10^{10}~\text{years}$ are out of the present scope, the $\omega$-dependent part
of \eqref{disp_mond} reduces to a numerically small factor.

Finally, we consider the $\sigma_0$-dependent part of
\eqref{disp_mond}. Consistent with the ``weak clustering hypothesis''
\eqref{WC1}--\eqref{WC2}, we are expecting the background density profile
$\sigma_0$ to be almost homogeneous. Thus, the factor $\partial_r
\sigma_0/\sigma_0$ will be of the order of the inverse of the characteristic
length scale $L$ of variation of $\sigma_0$ assumed to be far larger than the
typical size $\ell$ of the galaxy, which implies $\vert \bm{k} \cdot \bm{x}
\vert \lesssim \ell / L \simeq 0$ in \eqref{perturb}. A longitudinal
perturbation would therefore keep oscillating at the frequency $\omega$
without propagating, and we conclude that it would be stable.

\bibliography{/tmp_mnt/netapp/users_home4/letiec/Articles/ListeRef}

\end{document}